\documentclass[preprints,article,accept,moreauthors,pdftex]{Definitions/mdpi} 

\firstpage{1} 
\pubvolume{1}
\issuenum{1}
\articlenumber{0}
\pubyear{2021}
\copyrightyear{2021}
\history{Received: date; Accepted: date; Published: date}

\usepackage{amssymb}
\usepackage{ulem}

\Title{Fundamental Tone and Overtones
of Quasinormal Modes in Ringdown Gravitational Waves:
A Detailed Study in Black Hole Perturbation}

\Author{
Norichika Sago $^{1,2}$
\orcidA{} 
Soichiro Isoyama $^{3}$
\orcidB{}
Hiroyuki Nakano $^{4}$
\orcidC{}
}

\AuthorNames{Norichika Sago}
\AuthorNames{Soichiro Isoyama}
\AuthorNames{Hiroyuki Nakano}

\address{%
$^{1}$ \quad 
Department of Physics, Kyoto University, Kyoto 606-8502, Japan; \\
$^{2}$ \quad 
Advanced Mathematical Institute, Osaka City University, Osaka 558-8585, Japan; sago@tap.scphys.kyoto-u.ac.jp\\
$^{3}$ \quad 
School of Mathematics, University of Southampton, Southampton SO17 1BJ, United Kingdom; isoyama@yukawa.kyoto-u.ac.jp\\
$^{4}$ \quad 
Faculty of Law, Ryukoku University, Kyoto 612-8577, Japan;\\ 
~~~\quad hinakano@law.ryukoku.ac.jp\\
}

\corres{Correspondence: sago@tap.scphys.kyoto-u.ac.jp}

\abstract{
Ringdown gravitational waves 
of compact object binaries observed by ground-based 
gravitational-wave detectors encapsulate 
rich information to understand remnant objects 
after the merger
and to test general relativity in the strong field.
In this work, we investigate the ringdown gravitational waves in detail 
to better understand their property,
assuming that the remnant objects are black holes. 
For this purpose, we perform numerical simulations of 
post-merger phase of binary black holes by using the black hole 
perturbation scheme with the initial data given under the close-limit 
approximation, and generate data of ringdown gravitational waves
with smaller numerical errors than that associated with currently 
available numerical relativity simulations.
Based on the analysis of the data, we propose an orthonormalization 
of the quasinormal mode functions describing the fundamental tone and 
overtones to model ringdown gravitational waves.
Finally, through some demonstrations of the proposed model, 
we briefly discuss the prospects for ringdown gravitational-wave 
data analysis including the overtones of quasinormal modes.
}


\begin{document}

\section{Introduction}
\label{sec:introduction}

The remnant of merging black holes (BHs) is
a perturbed compact object 
(a Kerr BH~\cite{1963PhRvL..11..237K},
or even something more exotic)
characterized by the set of complex frequencies known as  
quasinormal modes (QNMs), 
and the gravitational radiation from this remnant is called 
the ringdown phase of gravitational waves (GWs). 
Ringdown GWs from binary BH mergers 
are now routinely detected by ground based GW detectors 
such as Advanced LIGO and Virgo~\cite{LIGOScientific:2020ibl,LIGOScientific:2021usb}. 
An accurate measurement of QNMs encoded 
in the ringdown signal therefore 
offers various GR tests regarding the compact object, 
for example,  
to disclose the remnant property  
(the ergo region of Kerr 
geometry~\cite{Nakamura:2016gri,Nakamura:2016yjl,Nakano:2016sgf}, 
ringdown test of general relativity (GR)~\cite{Nakano:2015uja} etc.), 
and to verify GR itself in the strong-field regime; 
we refer readers to Ref.~\cite{Berti:2009kk} 
for a review of the BH QNMs, 
and Ref.~\cite{Berti:2018vdi} for that of the ringdown GWs. 

The theoretical investigation of QNMs has a long history 
since early 1970, 
including Vishveshwara~\cite{Vishveshwara:1970zz},
Press~\cite{Press:1971wr}, Teukolsky and Press~\cite{Teukolsky:1974yv}, 
and Chandrasekhar and Detweiler~\cite{Chandrasekhar:1975zza}, and Detweiler~\cite{Detweiler:1977gy}.
In 1980s, Detweiler~\cite{Detweiler:1980gk} basically completed
the analysis of QNMs for Kerr BHs. 
Leaver~\cite{Leaver:1985ax} gave a standard method
to calculate the QNM frequencies very accurately 
(see also, e.g., Refs.~\cite{Castro:2013lba,Casals:2019vdb,Hatsuda:2020egs} 
for more modern techniques to compute QNMs, 
motivated by the recent development 
in the BH perturbation theory or high-energy physics). 

At the same time, the work on the data analysis 
of ringdown GWs
also goes back to 1980s. 
Echeverria~\cite{Echeverria:1989hg}
and Finn~\cite{Finn:1992wt} showed that
one can indeed extract information of the mass
and spin of the remnant Kerr BH from the ringdown signals.
Flanagan and Hughes~\cite{Flanagan:1997sx,Flanagan:1997kp} 
evaluated the signal-to-noise ratio (SNR) and parameter
estimation errors for practical GW detectors.
Creighton~\cite{Creighton:1999pm} analyzed data 
from the Caltech 40-meter prototype interferometer 
with a single-filter search.
Arnaud et al.~\cite{Arnaud:2002kf}
and Nakano et al.~\cite{Nakano:2003ma,Nakano:2004ib}
developed effective search methods of ringdown GWs.
Tsunesada et al.~\cite{Tsunesada:2004tc,Tsunesada:2004ft,Tsunesada:2005fe} 
discussed the detection efficiency, event selection, 
and parameter estimation of TAMA 300 (`prototype' KAGRA) data~\cite{TAMA:2001mcr}
in the matched filtering analysis.
The LIGO-Virgo data in 2005
and the LIGO-Virgo data in 2005--2010 were analyzed 
in Refs.~\cite{LIGOScientific:2009diy,LIGOScientific:2014wvq},
respectively
(see also Refs.~\cite{Goggin:2008dz,Caudill:2011kv,Caudill:2012qfa,Talukder:2012jga,Baker:2013phd} for the earlier related works).

Among that orderly development, the first direct GW detection of
merging binary BHs: GW150914~\cite{LIGOScientific:2016aoc} 
has particularly burst the activity 
of QNMs and the ringdown GW data analysis. 
In Ref.~\cite{LIGOScientific:2016lio}, 
three tests of GR involving the ringdown phase were attempted for GW150914:  
consistency test between the remnant BH parameters
evaluated by the inspiral and post-inspiral signals; 
consistency check between the remnant BH parameters
predicted from an inspiral-merger-ringdown waveform in GR
and evaluated by the least-damped QNM;  
and constraining deviations from GR
with a parametrized waveform model.
This analysis was based on the leading order, longest-lived QNM, 
i.e., the $(\ell=2,\,m=2,\,n=0)$ mode; 
here $(\ell,\,m)$ are angular indices, and $n$ shows the overtone index. 
(From this point onward we refer to $n=0$ mode as the fundamental tone,
while $n \geq 1$ mode as the overtone),
and it was a mainstream strategy for the ringdown GW data analysis 
(see, e.g., Ref.~\cite{Nakano:2021bbw} in the context of multiband GW observation 
with B-DECIGO~\cite{Nair:2015bga,Nakamura:2016hna,Isoyama:2018rjb,Kawamura:2020pcg}).

Nevertheless, Berti et al.~\cite{Berti:2007zu} showed that 
the ringdown analysis with only $(\ell=2,\,m=2,\,n=0)$ mode 
can lose $10\%$ of potential LIGO events, 
and indeed the recent development has revealed the importance 
of including QNM overtones in the ringdown GW analysis. 
Notably, Giesler et al.~\cite{Giesler:2019uxc} demonstrated 
that one can have a larger SNR for the ringdown phase
and better parameter estimations of the remnant BH
with enough QNM tones up to $n = 7$
(see also Refs.~\cite{Buonanno:2006ui,Baibhav:2017jhs,Carullo:2018sfu} 
for earlier work). 
The study was followed up with more detailed theoretical 
investigation~\cite{Bhagwat:2019dtm,Ota:2019bzl,Cook:2020otn,JimenezForteza:2020cve,Mitman:2020pbt,Dhani:2020nik,Forteza:2021wfq,Dhani:2021vac} 
to consider the QNM fits including overtones 
to the numerical relativity (NR) waveforms 
of binary BH mergers~\cite{Pretorius:2005gq,Campanelli:2005dd,Baker:2005vv} 
after the time of peak amplitude (of some waveform quantity). 
The observational impacts of including QNM overtones in the ringdown GW analysis 
are examined for GW150914~\cite{Isi:2019aib} and 
for GW190521~\cite{LIGOScientific:2020iuh}, 
but the strong contribution of overtones and higher harmonics
was not found in the ringdown signal. 
Because GW190521 with a total mass of $150\,M_{\odot}$ is 
the heaviest binary BH merger observed to date 
and the its signal are dominated 
by the merger and ringdown phases, 
three ringdown models were 
particularly considered in this case~\cite{LIGOScientific:2020ufj}: 
a single damped sinusoid, 
a set of the $(\ell=2,\,m=2)$ QNM fundamental tone and 
overtones, i.e., $n=0,\,1,\,2$, and 
a set of all fundamental tone of the QNMs 
up to the $\ell=4$ harmonic mode
by taking the QNM starting time into account
(see also Ref.~\cite{LIGOScientific:2020tif} 
about testing GR with ringdown GWs
for binary BH events in GWTC-2~\cite{LIGOScientific:2020ibl}).

To data, the contribution of QNM overtones to the ringdown 
has been largely investigated relying on the NR waveforms 
as a reference signal. 
Although NR waveforms `exactly' describe the ringdown phase of the BH mergers, still, 
the relative error in the NR strain data would be typically limited at $\sim 10^{-4}$
around the time of peak amplitude 
(see Figure~2 of Ref.~\cite{Giesler:2019uxc}),
according to a publicly available NR catalogue 
of ``SXS Gravitational Waveform Database''~\cite{SXS}; 
see also other NR catalogues including 
``CCRG@RIT Catalog of Numerical Simulations''~\cite{RIT},
``Georgia tech catalog of gravitational waveforms''~\cite{Gatech} 
and 
``SACRA Gravitational Waveform Data Bank''~\cite{SACRA}. 

Interestingly, it is found that the dominant $(\ell=2,\,m=2)$ mode 
after the time of the peak amplitude could be fitted by only the QNM fundamental tone and overtones 
in the linearized perturbation around the remnant BH spacetime, 
making use of precise 
NR waveforms~\cite{Okounkova:2020vwu,Pook-Kolb:2020jlr,Mourier:2020mwa}. 
This facts implies that the second and higher order QNMs would not be excited 
enough at least in the $(\ell=2,\,m=2)$ mode 
(we note, however, that the second order QNMs 
due to the self-coupling of the first-order, 
e.g.,  $(\ell=2,\,m=2)$ mode was already identified 
in the $(\ell=4,\,m=4)$ mode of the NR waveforms~\cite{London:2014cma}; 
see also, e.g., Refs.~\cite{Ioka:2007ak,Nakano:2007cj,Okuzumi:2008ej} 
for earlier works on the second order QNMs).
This motivates us to work with the close-limit approximation 
to the binary BH merges, proposed in the seminal work 
by Price and Pullin~\cite{Price:1994pm}, and Abrahams and Prices~\cite{Abrahams:1995wd}. 

The close-limit approximation describes the last stage 
of a binary BH coalescence 
as a perturbation of a single (remnant) BH, 
demonstrating a remarkably good agreement 
with full NR simulation~\cite{Anninos:1998wt,Sopuerta:2006wj}; 
other recent applications of close-limit approximations are, for example, 
for the initial data of the NR simulation~\cite{Baker:2001sf} 
and for binary BH mergers beyond GR~\cite{Annulli:2021dkw}. 
A particularly relevant aspect of the closed-limit approximation here 
is that the approximated system can be evolved highly accurately 
in the framework of the linear BH perturbation theory. 
A range of methods for numerically evolving the linearized Einstein equation 
is described in literature, and the numerical accuracy 
that we can achieve in this work is at the level of $10^{-7}$ 
in the GW strain data at the time of the peak amplitude
and $10^{-13}$ in the late time of simulations
(see Figure~\ref{fig:code-convergence}).
Therefore, our close-limit waveforms 
in the BH perturbation approach allow us 
the more depth study regarding to the QNM fit 
than using NR waveform 
with numerically accuracy available for now. 

This paper is organized as follows.
In Section~\ref{sec:CL-metric}, 
we briefly review the close-limit approximation of the 
post-Newtonian (PN) metric~\cite{LeTiec:2009yf}, which we use to
prepare the initial data for the BH perturbation calculation.
In the calculation presented in this work, for simplicity, we focus 
on the head-on collision of nonspinning binary BHs.
Therefore, the remnant BH can be considered 
as a (nonspinning) Schwarzschild BH.
Our numerical method in the BH perturbation approach
is presented in Section~\ref{sec:method}
where we carefully check the numerical accuracy. 
In Section~\ref{sec:QNM}, we 
propose a modified QNM fitting formula by introducing an 
orthonormal set of mode function to analyze the ringdown waveforms.
In Section~\ref{sec:results}, we use the QNM fitting formulae 
in Section~\ref{sec:QNM} and apply them 
to the GW data obtained in Section~\ref{sec:method}.
We will find the late-time power-law tail~\cite{Price:1971fb} 
in the fit residuals, 
and a convergent behavior 
for the modified QNM fitting formula.
In Section~\ref{sec:discussion}, 
we summarize our analysis, and discuss 
some application.
A brief analysis of 
the dominant $(\ell=2,\,m=2)$ mode of NR waveforms 
for nonprecessing binary BHs
and the fitting 
with the QNMs overtones is given in 
Appendices~\ref{app:fit_NR} and~\ref{app:fit-coeffs-NR}.
Here, we find some problematic behavior 
in NR waveforms for the cases
with a highly spinning remnant BH, 
and this is discussed in Appendix~\ref{app:qnm_freq}.

Throughout this paper, 
we adopt the negative metric signature $(-+++)$, geometrized 
units with $c=G=1$, except in Section~\ref{sec:CL-metric}.
The coordinates, $x^{\alpha} = \{ ct,\,r,\,\theta,\,\varphi \}$,
are used as the Schwarzschild coordinates for Schwarzschild spacetime
and the Boyer-Lindquist coordinates for Kerr spacetime.

\section{Post-Newtonian initial conditions in the close limit: 
head-on collisions}
\label{sec:CL-metric}

We wish to describe the evolution of nonspinning binary BHs 
in its last stage, 
making use of the close-limit approximation starting from a 
given initial condition. 
For this purpose, we adopt the close-limit form of the post-Newtonian metric 
for two nonspinning BHs (or point particles) 
of masses $m_A\,(A = 1,\,2)$ derived 
by Le Tiec and Blanchet~\cite{LeTiec:2009yf}.
This close-limit, PN metric is particularly convenient 
to examine 
the ringdown phase of asymmetric mass-ratio binaries.
To set the stage for our analysis, we review their essential discussion 
and results here.

We assume that the binary system admits two different 
(dimensionless) small parameters. 
The first one is the close-limit parameter~\cite{Price:1994pm,Abrahams:1995wd,Gleiser:1996yc,Andrade:1996pc,Gleiser:1998rw,Sopuerta:2006wj} 
in which the binary separation $r_{12}$ is sufficiently 
small compared to the distance $r$ from any field point to 
a reference source point (e.g., the center of mass of the system): 
\begin{equation} \label{eq:eps_CL}
    \epsilon_{\rm CL} \sim \frac{r_{12}}{r} 
    \ll 1 \,.
\end{equation}
The other is the post-Newtonian (PN) 
parameter~\cite{Futamase:2007zz,Blanchet:2013haa,Levi:2018nxp,Schafer:2018kuf} 
where the typical value of the relative orbital velocity $v_{12}$ 
is small enough compared to the speed of light 
(or that of the separation between two masses $r_{12}$ is sufficiently 
large):  
\begin{equation}\label{eq:eps_PN}
    \epsilon_{\rm PN} \sim \frac{v_{12}^2}{c^2}
    \sim \frac{G M}{c^2 r_{12}} \ll 1 \,.
\end{equation}
Here, $M = m_1 + m_2$ is the total mass of the binary system, 
and we shall say that a term relative to $O(\epsilon_{\rm PN}^N)$ 
is $N$th PN order. We note that the PN expansion is valid 
when $r \ll r_{12} / \sqrt{G M / (c^2 r_{12})}$ 
that defines the (so named) near-zone. 

The crux of this approach is that the metric of the binary system is 
{formally} expanded in powers of both in $\epsilon_{\rm CL}$ 
and $\epsilon_{\rm PN}$ 
(though the close-limit approximation would imply 
$r_{12} \gtrsim G M /c^2$ i.e., $\epsilon_{\rm PN} \lesssim 1$~\cite{Price:1971fb}; 
see also Refs.~\cite{LeTiec:2009yf,LeTiec:2009yg,Nichols:2010qi} 
for an elaboration of this double expansion). 
This allows us to recast the (late inspiral phase of) PN binary spacetime 
to a PN vacuum perturbation of the single Schwarzschild spacetime. 
Specifically, we work with the $2$PN near-zone metric $g_{\mu \nu}^{\rm PN}$
of two point masses~\cite{Blanchet:1998vx, Tagoshi:2000zg, Faye:2006gx,Bohe:2012mr}, and then further expand it 
in the power of $\epsilon_{\rm CL}$. 
The resultant metric can be manipulated to be identified with 
\begin{equation}
g_{\mu \nu}^{\rm PN} = g_{\mu \nu}^{\rm Schw.} + h_{\mu \nu}\,,
\end{equation}
where $g_{\mu \nu}^{\rm Schw.}$ is the Schwarzschild metric with mass $M$ 
and $h_{\mu \nu}$ is the perturbation that takes the schematic form
\begin{equation}\label{h-CL}
h_{\mu \nu} = 
G\, \sum_{n = 0}^2 \,\sum_{k \geq 0}\, 
h_{\mu \nu}^{(n,\,k)}\, \epsilon_{\rm PN}^n\, \epsilon_{\rm CL}^{k + 1}
+ O(G^2)\,.
\end{equation}
We note that this expression is truncated at $O(G)$ 
so that $h_{\mu \nu}$ consistently satisfies the linearized 
Einstein equation to the $2$PN order. 
The general explicit expressions for the coefficients $h_{\mu \nu}^{(n,\,k)}$ 
are too unwieldy to be presented here; the expressions up to $k = 2$  
in the (slightly altered) Cartesian harmonic coordinates are listed 
in Eq.~(2.12) of Ref.~\cite{LeTiec:2009yf}, for example.

Because the Schwarzschild spacetime is spherically symmetric, 
the angular dependence of $h_{\mu \nu}$ in Eq.~\eqref{h-CL} 
can be conveniently separated to decompose it 
in a given tensor spherical harmonics. 
Using a standard Schwarzschild coordinates, 
this mode decomposition takes a form 
(e.g., Section~4.2 of Ref.~\cite{Pound:2021qin}) 
\begin{equation}
h_{\mu\nu} 
=
\sum_{i=1}^{10} \sum_{\ell,\,m} 
{\bar h}^{(i),\,\ell m}(t,\,r) 
Y^{(i),\,\ell m}_{\mu\nu}(\theta,\,\varphi)\,,
\end{equation}
where $Y^{(i),\,\ell m}_{\mu\nu}$ is a basis of 
tensor spherical harmonics. 
A set of 10 time-radial functions ${\bar h}^{(i),\,\ell m}$ 
further decouples into two subsets: 
seven for even-parity perturbations while three for odd-parity ones, 
depending on whether they are ``even'' or ``odd'' 
under the transformation $(\theta,\,\varphi) \to (\pi - \theta,\,\pi + \varphi)$.

\subsection{Model problem: head-on collisions of two nonspinning black holes}
\label{subsec:head-on}

We now specialize the close-limit, $2$PN metric to a head-on collision 
of two nonspinning BHs. 
The head-on collision is not an astrophysically relevant scenario, 
but it provides a useful and simple example to explore 
the ringdown radiation with a good numerical accuracy
in the remaining sections. 

We adopt the notations of Le Tiec and Blanchet 
(Section~3 of Ref.~\cite{LeTiec:2009yf}), 
and work with the Regge-Wheeler basis of 
tensor Spherical harmonics and in the Regge-Wheeler gauge 
to simplify the expressions~\cite{Regge:1957td} 
(see also Appendix~A of Ref.~\cite{Sago:2002fe}) . 
In the case of the head-on collision, the odd-parity perturbations are 
identically zero, and the non-trivial components of even-parity perturbations 
(in the Regge-Wheeler gauge) are 
\begin{align}
h_{00} &= \left( 1 - \frac{2M}{r} \right) 
\sum_{\ell,\,m} {\tilde H}^{\ell m}_0\, Y_{\ell m}\,, \\
h_{rr} &= \left( 1 - \frac{2M}{r} \right)^{-1} 
\sum_{\ell,\,m} {\tilde H}^{\ell m}_2\, Y_{\ell m}\,, \\
{h_{\varphi \varphi}} &= 
h_{\theta \theta}\,{\sin^2 \theta} = r^2 
\sum_{\ell,\,m} {\tilde K}^{\ell m}\,{\sin^2 \theta}\, Y_{\ell m}\,.
\end{align}
We then read off $ {\tilde H}^{\ell m}_0,\,{\tilde H}^{\ell m}_2$ 
and ${\tilde K}^{\ell m}$ from the close-limit expansion to the $2$PN metric 
in Eq.~\eqref{h-CL}. 
To model the ringdown radiation yet in the simple setup, we shall consider 
only the $\ell = 2$ and $3$ perturbations that satisfy the linearized Einstein 
equations at the first order in the (symmetric) mass ratio 
$\nu \equiv m_1 m_2 / M^2$ up to terms $O(G^2,\,c^{^6},\,r_{12}^4)$. 
This restriction gives
\begin{equation}
{\tilde H}^{\ell m}_{0}
\equiv
{\tilde H}^{\ell m}
= 
{\tilde H}^{\ell m}_{2}\,.
\end{equation}
With this relation, the explicit expressions of $\ell = 2$ components on the time symmetric initial surface are 
(Eq.~(3.6) of Ref.~\cite{LeTiec:2009yf}; $r_{12}$ is the binary's separation 
and we assume that the relative orbital velocity $v_{12}$, 
orbital angular velocity $\omega_{12}$ and the orbital phase $\beta$ 
are all zero) 
\begin{align}\label{RWZ-ell2}
\tilde{H}_\textrm{CL}^{2,0} 
=& 
-2\sqrt{\frac{\pi}{5}}\, \nu\, \frac{M\,r_{12}^2}{r^3} 
=
\tilde{K}_\textrm{CL}^{2,0}\,, \cr
\tilde{H}_\textrm{CL}^{2,\pm2} 
=& -\sqrt{\frac{3}{2}}\,  \tilde{H}_\textrm{CL}^{2,0}\,,
\cr
\tilde{K}_\textrm{CL}^{2,\pm2}
=& -\sqrt{\frac{3}{2}}\,  \tilde{K}_\textrm{CL}^{2,0}\,,
\end{align}
and those of $\ell = 3$ components are 
\begin{align}\label{RWZ-ell3}
\tilde{H}_\textrm{CL}^{3,\pm 1} 
=&
\mp \sqrt{\frac{3\pi}{7}}\, \nu\, \frac{\delta M r_{12}^3}{r^4}
= 
\tilde{K}_\textrm{CL}^{3,\pm 1}\,, \cr 
\tilde{H}_\textrm{CL}^{3,\pm3} 
=& 
-\sqrt{\frac{5}{3}}\, \tilde{H}_\textrm{CL}^{3,\pm1}\,,
\cr
\tilde{K}_\textrm{CL}^{3,\pm3}
=&
-\sqrt{\frac{5}{3}}\, \tilde{K}_\textrm{CL}^{3,\pm1}\,,
\end{align}
with $\delta M \equiv m_1 - m_2$.

\section{Numerical method}
\label{sec:method}

In this section,
we shall use the close-limit, $2$PN metric in Eqs.~\eqref{RWZ-ell2} and~\eqref{RWZ-ell3} 
(in the Regge-Wheeler gauge) as initial data, 
and numerically evolve it with the time-domain (TD) implementation 
of the Regge-Wheeler-Zerilli equations, 
to obtain the associated ringdown waveform. 

Before proceeding, we note that Eq.~\eqref{RWZ-ell2} 
in the equal-mass limit $(\nu = 1/4)$ essentially 
agrees with the Brill-Lindquist geometry 
for the head-on collision~\cite{Brill:1963yv},  
after a certain coordinate adjustment to the initial separation $r_{12}$. 
This will provide reassurance that the physics in our TD numerical waveform 
should be largely independent of our specific choice 
of the PN initial data
(see Section~5 of Ref.~\cite{LeTiec:2009yf} 
for further elaboration).

\subsection{Initial data and gravitational waveform
in terms of master functions}
\label{subsec:init}

In the practical implementation, it is more preferable 
to use gauge-invariant master functions that can be constructed from 
$\tilde{H}_\textrm{CL}^{\ell m}$ and $\tilde{K}_\textrm{CL}^{\ell m}$
in the Regge-Wheeler gauge. 
For even-parity perturbations, 
we use (the Regge-Wheeler-gauge expression for) 
the Zerilli-Moncrief function   
given by (Eq.~(5.1) of Ref.~\cite{LeTiec:2009yf}; 
see also Eq.~(126) of Ref.~\cite{Pound:2021qin}) 
\begin{equation}\label{def-ZMfunc}
\Psi^{\ell m} =
\frac{r}{2(\lambda_\ell + 1)} \left[
\tilde{K}^{\ell m}
+ \frac{r - 2M}{\lambda_\ell\, r + 3M} 
\left(
\tilde{H}^{\ell m} - r\, \partial_r \tilde{K}^{\ell m}
\right) \right]\,,
\end{equation}
where $\lambda_\ell \equiv  (\ell - 1) (\ell + 2) / 2$. 
Substituting Eqs.~\eqref{RWZ-ell2} and~\eqref{RWZ-ell3} into this, 
the $\ell = 2$ and $3$ multipolar coefficients for the head-on collisions 
are then translated to 
\begin{align}
\Psi^{2,\,0}
&= \frac{1}{3} \sqrt{\frac{\pi}{5}}\,  \nu M 
\frac{r_{12}^2}{r^2} \frac{5 M - 6 r}{3 M + 2 r}\,, 
\label{eq:IC-ZM}
\\
\Psi^{3,\,\pm1} 
&= \pm \frac{1}{4} \sqrt{\frac{\pi}{21}}\, \nu \delta M 
\frac{r_{12}^3}{r^3} \frac{7 M - 10 r}{3 M + 5 r}\,,
\label{eq:IC-ZM3}
\end{align}
with $\Psi^{2,\,\pm2} = -\sqrt{{3}/{2}} \, \Psi^{2,\,0}$ 
and $\Psi^{3,\,\pm3} = -\sqrt{{5}/{3}} \,  \Psi^{3,\,\pm1}$. 

Furthermore, the Zerilli-Moncrief function in Eq.~\eqref{def-ZMfunc} 
is directly related to the GW strain 
\begin{equation}\label{eq:RWZwaves}
h_{+} - ih_{\times} =
\sum_{\ell, m} 
\frac{\sqrt{(\ell-1)\ell(\ell+1)(\ell+2)}}{r} 
\Psi^{\ell m} (t,r) \, {}_{-2}Y_{\ell m}(\theta, \varphi)\,,
\end{equation}
where ${}_{-2}Y_{\ell m}$ is the spin-weighted spherical harmonics 
with $s = -2$ (note that the contribution from the odd-parity
perturbation identically vanishes in the case of the head-on collision).
To avoid the dependence of the polarization on the observer's location, 
we focus on $\Psi^{\ell m}$ in the remaining sections.

\subsection{Time domain integration of the Zerilli equation}
\label{subsec:TDcode}

In our model problem of the head-on collision, 
the Zerilli-Moncrief function $\Psi^{\ell m}$ satisfies the homogeneous 
Zerilli equation. 
We write the equation in the double null coordinates, in practice,
\begin{equation}\label{eq:RWZ-eq-null}
\left[\frac{\partial^2}{{\partial u \partial v}} 
+ \frac{1}{4} V_\ell^{\rm Z}(r) \right]
\Psi^{\ell m}(u,v) = 0 \,,
\end{equation}
where $u=t-r^*$ and $v=t+r^*$ with the tortoise coordinate 
$r^*=r+2M\ln[r/(2M)-1]$, and 
\begin{equation}
V_\ell^{\rm Z}
=
\left(1 - \frac{2 M}{r} \right)
\left[
\frac{\ell (\ell + 1)}{r^2}
- \frac{6M}{r^3}
\frac{\lambda_\ell (\lambda_\ell + 2)\, r^2 + 3 M (r - M)}
{(\lambda_\ell\, r + 3 M)^2}
\right] \,,
\end{equation}
is the Zerilli potential. 

\begin{figure}[ht]
\begin{tabular}{cc}
  \begin{minipage}[t]{0.55\hsize}
    \centering
    \includegraphics[width=0.8\hsize]{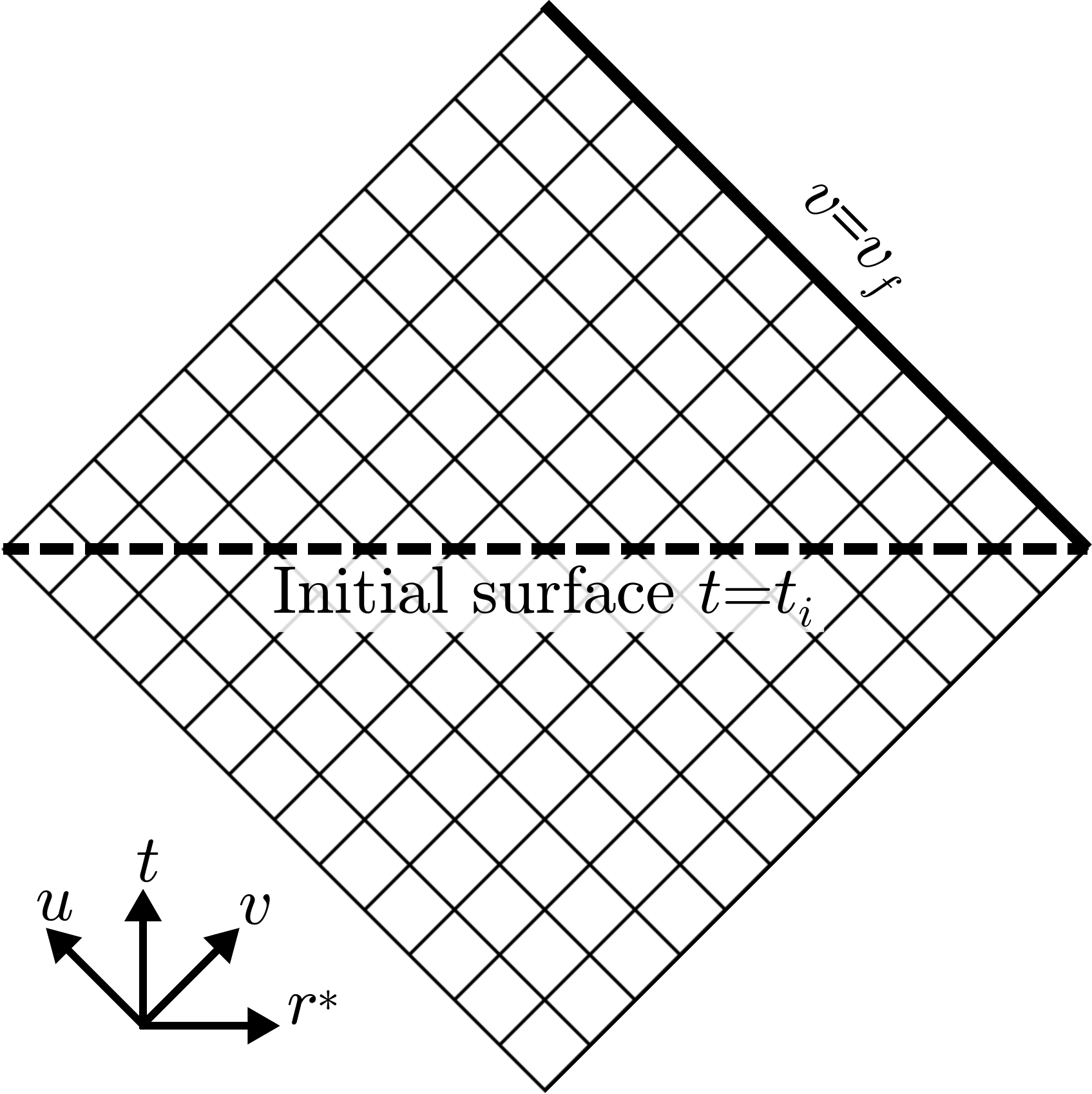}
  \end{minipage} &
  \begin{minipage}[t]{0.35\hsize}
    \centering
    \includegraphics[width=0.8\hsize, bb = 0 -60 227 239]{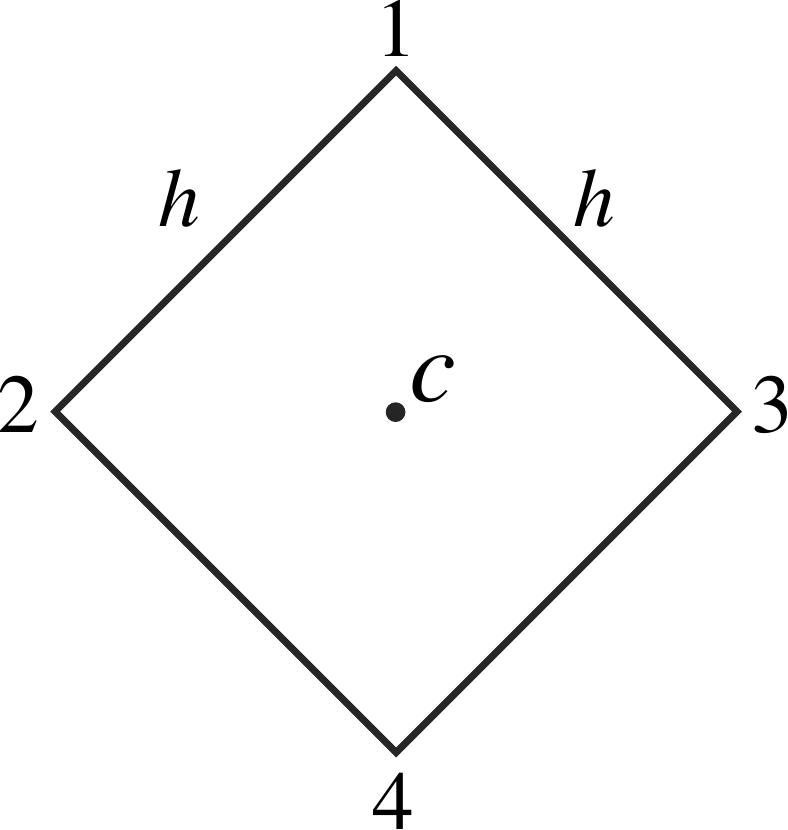}
  \end{minipage}
\end{tabular}
\caption{\textit{Left}: Schematic picture of the numerical domain in
the double null coordinates,
$(u=t-r^*,\,v=t+r^*)$. We impose the close-limit, PN initial condition 
in Eqs.~\eqref{eq:IC-ZM} and~\eqref{eq:IC-ZM3}
on the surface with $t=t_i$ (the bold dashed line)
and extract the waveform on the surface with $v=v_f$ (the bold solid line).
\textit{Right}: a cell in the numerical domain with the size of $h \times h$, 
where $h$ is the resolution and $C$ denotes the center of the cell.}
\label{fig:domain-cell}
\end{figure}

To evolve Eq.~\eqref{eq:RWZ-eq-null}, we use the finite-difference scheme
developed by Lousto and Price~\cite{Lousto:1997wf}.
We introduce an uniform null grid in the $1 + 1$ numerical domain 
as the left panel of Figure~\ref{fig:domain-cell}.
Consider a grid cell with the size of $h \times h$ in the domain. 
Integrating Eq.~\eqref{eq:RWZ-eq-null} over the cell gives
\begin{equation}
\Psi_1 - \Psi_2 - \Psi_3 + \Psi_4
+ \frac{h^2}{8}V_\ell^\textrm{Z}(r_c)(\Psi_2 + \Psi_3) + O(h^4) = 0 \,,
\end{equation}
where $r_c$ is the value of $r$ at the center of the cell, 
and $\Psi_i$ for $i=1,\,2,\,3,\,4$ correspond to the values of $\Psi$ 
(we have omitted the indices $(\ell,\,m)$ here)
at the vertices of the cell in the right panel of Figure~\ref{fig:domain-cell}.
By using the above equation,
we can obtain $\Psi_1$ from $\Psi_2$, $\Psi_3$ and $\Psi_4$
with the local error term of $O(h^4)$,
\begin{equation}
\Psi_1 = \Psi_2 + \Psi_3 - \Psi_4 
- \frac{h^2}{8}V_\ell^\textrm{Z}(r_c)(\Psi_2 + \Psi_3) + O(h^4) \,.
\label{eq:FDS-vacuum}
\end{equation}
The local error is accumulated through the integration over $v$ and $u$
to give a global error of $O(h^2)$. 

The finite-difference scheme on the initial surface should be derived separately.
To implement the time-symmetric initial conditions in Eqs.~\eqref{eq:IC-ZM} and~\eqref{eq:IC-ZM3},
we also assume the time symmetry on the initial surface,
and therefore $\partial_t \Psi|_{t=t_0} = 0$, which leads to
\begin{equation}
\Psi_1 = \Psi_4 + O(h^3) \,,
\end{equation}
if the vertices 2 and 3 are on the surface.
From this relation, we find that the finite-difference 
scheme on the initial surface is 
\begin{equation}
\Psi_1 = \frac{\Psi_2 + \Psi_3}{2}
- \frac{h^2}{16}V_\ell^\textrm{Z}(r_c)(\Psi_2 + \Psi_3) + O(h^3) \,.
\label{eq:FDS-initial}
\end{equation}
While the local error term in this case is $O(h^3)$, 
it eventually gives the global error of $O(h^2)$ 
because the error is accumulated only over 
the initial surface.
Hence, the convergence of our numerical solution 
through the finite difference scheme in
Eqs.~\eqref{eq:FDS-vacuum} and~\eqref{eq:FDS-initial} 
is quadratic with respect to $h$ 
(see Figure~\ref{fig:code-convergence} below).

\subsection{Simulation parameters}
\label{subsec:SP}

In this paper, we calculate the Zerilli-Moncrief function with the
close-limit, $2$PN initial data for head-on collisions of two 
nonspinning BHs in the following two cases:
\begin{itemize}
\item Case A: 
an equal mass collision with the initial separation of $r_{12}=3.5M$.
\item Case B: 
an asymmetric mass collision with the mass ratio of $\nu=0.2$ 
and the initial separation of $r_{12}=4.4M$.
\end{itemize}
These system parameters are chosen so that they agree with 
those employed in Ref.~\cite{LeTiec:2009yf}. 
In the case A, we focus only on $(\ell = 2,\,m = 0)$ mode 
because $(\ell = 2,\,m = \pm2)$ modes are just proportional to this mode. 
Similarly, in the case B, we look at only $(\ell = 3,\,m = 1)$ modes 
because $(\ell = 3,\,m = -1)$ differs by only an overall minus sign, 
and $(\ell = 3,\,m = \pm3)$ modes are proportional to that mode; 
recall Eqs.~\eqref{eq:IC-ZM} and~\eqref{eq:IC-ZM3}.

\subsection{Code validation}

In our numerical simulations, 
the initial data given in Section~\ref{subsec:init} 
with the parameters shown in Section~\ref{subsec:SP} is set 
on the region of $-1000\le r^*/M \le 3000$ at $t=t_i=0$
and the Zerilli equation in Eq.~\eqref{eq:RWZ-eq-null} 
are integrated within 
the future domain of dependence of
the initial surface (the upper triangular
domain in the left panel of Figure~\ref{fig:domain-cell}).
The produced gravitational waveform is extracted on $v=v_f=3000M$ 
as a function of the retarded time $u=t-r^*$.
We take the resolution as $h/M=(0.08,\, 0.04,\, 0.02,\,
0.01,\, 0.005)$ with
the number of grid-point $N=(50000,\, 100000,\, 
200000,\, 400000,\, 800000)$ respectively.
From this point onward, we may set $M = 1$. 

\begin{figure}[ht]
\centering
\includegraphics[width=0.5\textwidth]{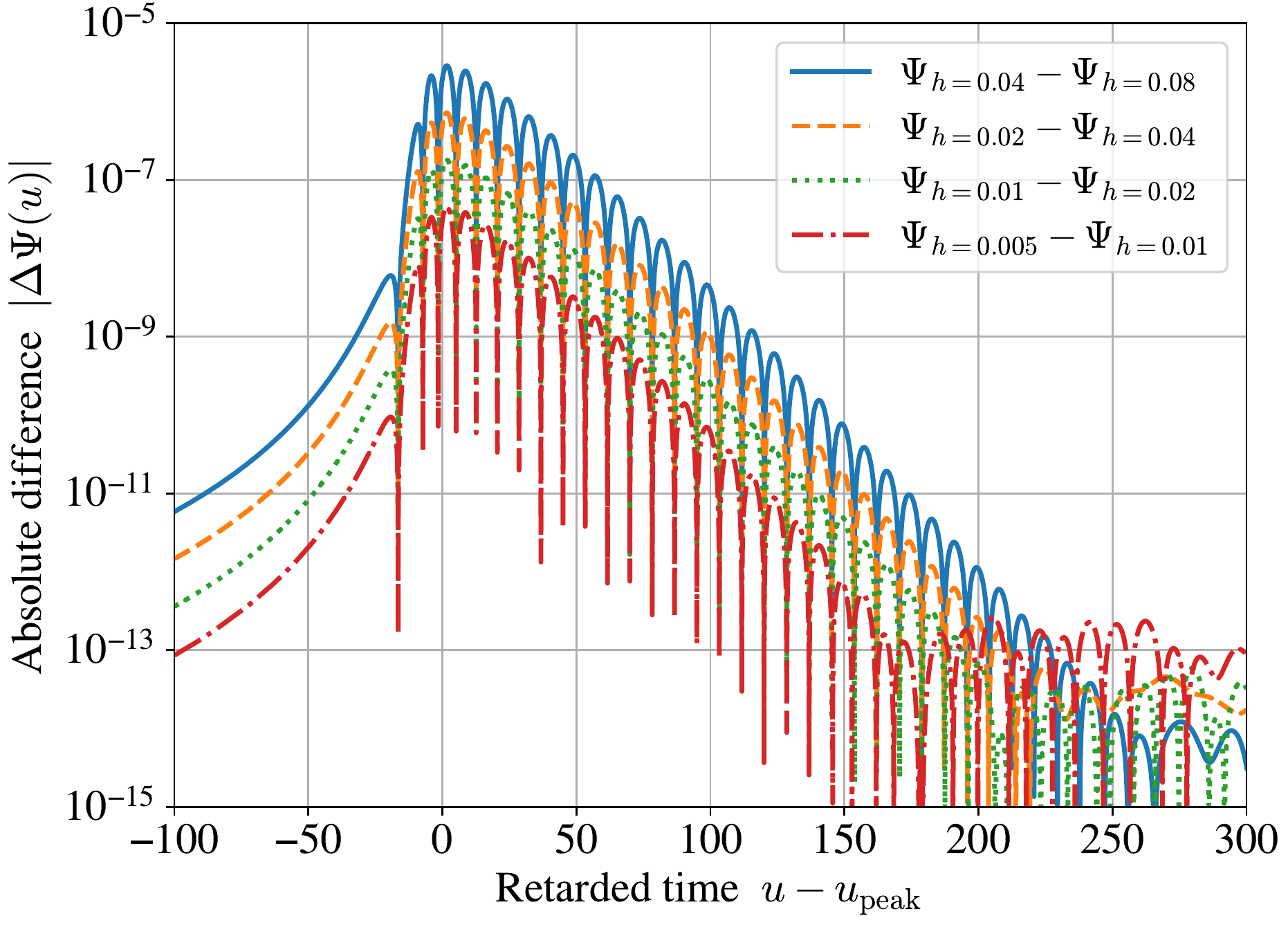}
\caption{
Absolute differences between the results with different resolutions,
$h/M=(0.08,\, 0.04,\, 0.02,\, 0.01,\, 0.005)$.
For example, $\Psi_{h=0.04}-\Psi_{h=0.08}$ (the blue solid curve)
denotes the absolute differences 
between $(\ell=2,\,m=0)$ mode of the Zerilli-Moncrief function 
for the case A in the $h/M=0.08$ and $0.04$ simulations. 
The plot demonstrates the second order convergence
of our time domain code.
The numerical accuracy with the highest resolution is roughly estimated 
as $10^{-7}$ at the time of the peak amplitude 
$u_{\rm peak}$ and $10^{-13}$
in the late time of $(u-u_{\rm peak})/M \gtrsim 200$.}
\label{fig:code-convergence}
\end{figure}

To check the convergence of our TD code, 
we consider the absolute differences between the Zerilli-Moncrief function 
$\Psi$ (we have omitted the indices $(\ell,\,m)$ here)
computed with different resolutions,
\begin{equation}
\Delta \Psi = \left|\Psi_{h_2}-\Psi_{h_1}\right|\,,
\label{eq:TDerror-estimator}
\end{equation}
where $h_1$ and $h_2$ denote adjacent resolutions.
The result for the $(\ell=2,\,m=0)$ Zerilli-Moncrief function
for the case A is shown in Figure~\ref{fig:code-convergence}.
We can find that the difference becomes a quarter when the
resolution gets a half. 
This shows the $O(h^2)$ convergence of the data, as expected. 
The numerical accuracy with the highest resolution, $h=0.005M$, 
is roughly estimated as $10^{-7}$ at the time of the peak amplitude.
In the late time of $(u-u_{\rm peak})/M \gtrsim 200$,
the numerical accuracy becomes $10^{-13}$.

\section{Modeling of ringdown waveforms}
\label{sec:QNM}

In this section, we introduce waveform models to describe 
the ringdown data in our close-limit calculations, 
as a superposition of QNMs associated with the remnant Schwarzschild BH.
The method to compute QNMs are well developed 
in literature (see, e.g., Ref.~\cite{QNM_EB}).
We use the accurate numerical data of QNM frequencies 
of the remnant BH provided by Berti~\cite{QNM_EB} 
(for $\ell=2$)
and those generated with the Black Hole Perturbation Club 
(B.H.P.C.) code~\cite{BHPC} (for $\ell=3$).

We should note that the zero-spin remnant BH here is 
a limitation of our close-limit approximation 
to the binary nonspinning BHs. 
When we treat general binary BH mergers 
in the full NR simulations with an astrophysically realistic initial data,
the remnant BHs after merger are Kerr with a non-zero spin 
in general. 
For example, a nonspinning equal-mass binary BH 
of the quasicircular merger can produce a remnant Kerr BH 
with the nondimensional spin $\sim 0.67$~\cite{Lousto:2009mf}.

\subsection{Standard quasinormal-mode fitting formula}
\label{subsec:standard-fit}

We write the ringdown waveforms as a sum of the fundamental tone 
and overtones of QNMs decomposed into spin-weighted spherical harmonics
with angular indices $(\ell,\,m)$ and spin $s = -2$,
${}_{-2}Y_{\ell m}(\theta, \varphi)$:
\begin{equation}
h = h_+ - ih_\times =
\frac{1}{r} \sum_{\ell, m} h_N^{\ell m}(t) 
{}_{-2}Y_{\ell m}(\theta, \varphi)
\quad (t \geq t_0)\,,
\label{eq:sperical-decomposition} 
\end{equation}
with 
\begin{equation}
h_N^{\ell m}(t) =
\sum_{n=0}^N Q_n^{\ell m} e^{-i\omega_{\ell m n}^{\rm QNM} (t-t_0)}\,,
\label{eq:complex-QNM-formula}
\end{equation}
where we truncate the summation with a finite overtone index 
of $n=N$ although there are infinite overtones.
Here, $Q_n^{\ell m}$ is a complex amplitude,
$\omega_{\ell m n}^{\rm QNM}$ is the Schwarzschild QNM frequency
of the $(\ell,\, m,\, n)$ mode,
and $t_0$ is the 'starting time' before which we do not include the model to fit our numerical waveforms. 
As mentioned later, we choose the time of the peak amplitude of the wave, 
$t_\textrm{peak}$, as $t_0$.

With an appropriate choice of the initial phase, 
our numerical waveforms of the head-on collisions 
are given as purely real functions of the retarded time, $u$.
Therefore, instead of the complex representation 
of the waveform in Eq.~\eqref{eq:complex-QNM-formula}, 
we adopt the following real expression 
for the Zerilli-Moncrief function in practice,  
making use of the relation in Eq.~\eqref{eq:RWZwaves},
\begin{equation}
\Psi_N^{\ell m}(u) =
\sum_{n=0}^N \left[ 
C_{2n}^{\ell m} \,\phi_{2n}^{\ell m}(u)
+ C_{2n+1}^{\ell m} \,\phi_{2n+1}^{\ell m}(u)
\right] \,, 
\label{eq:fit-formula-QNM}
\end{equation}
with
\begin{align}
\phi_{2n}^{\ell m}(u) &=
\sqrt{\frac{4(\omega_{\ell mn}^2\tau_{\ell mn}^2+1)}
           {\omega_{\ell mn}^2\tau_{\ell mn}^3}}
\,e^{-(u-u_0)/\tau_{\ell m n}}
\sin\left[ \omega_{\ell m n}(u-u_0) \right] \,,
\label{eq:fit-formula-QNM1}
\\
\phi_{2n+1}^{\ell m}(u) &=
\sqrt{\frac{4(\omega_{\ell mn}^2\tau_{\ell mn}^2+1)}
           {(\omega_{\ell mn}^2\tau_{\ell mn}^2+2)\tau_{\ell mn}}}
\,e^{-(u-u_0)/\tau_{\ell m n}}
\cos\left[ \omega_{\ell m n}(u-u_0) \right] \,,
\label{eq:fit-formula-QNM2}
\end{align}
where $C_k^{\ell m}$ ($k=0,\cdots,2N+1$) are real constants,
$u_0$ is the starting time in terms of the retarded time,
$\omega_{\ell mn}$ and $\tau_{\ell mn}$ are derived
from the real and imaginary parts of the QNM frequency,
respectively, 
\begin{align}
\omega_{\ell m n} = \Re (\omega_{\ell m n}^{\rm QNM}) \,,
\\
\tau_{\ell m n} = - \dfrac{1}
{\Im (\omega_{\ell m n}^{\rm QNM})} \,.
\label{eq:def-tau}
\end{align}
The prefactor in the mode functions 
$\phi_k^{\ell m}(u)$ of Eqs.~\eqref{eq:fit-formula-QNM1}
and~\eqref{eq:fit-formula-QNM2} is chosen 
so that they are certainly normalized (see the next subsection).
We note that the model in Eq.~\eqref{eq:fit-formula-QNM} 
is a linear function  
with respect to the fitting coefficients $C_{k}^{\ell m}$. 
This means that one can obtain the best fit values of $C_k^{\ell m}$ 
through linear least squares.

\subsection{Modified quasinormal-mode fitting formula with an orthonormal
set of mode functions}
\label{subsec:ON-fit}

The mode functions $\phi_k^{\ell m}(u)$ defined in the previous 
subsection are linearly independent but not orthogonal each other. 
For convenience of our analysis, 
we introduce an `orthonormal set' of mode functions 
by using the Gram-Schmidt procedure.
For this purpose, we first define the inner product between two real functions 
of $f(u)$ and $g(u)$ given in $u\ge u_0$ 
\begin{equation}
(f,\, g) \equiv
\int_{u_0}^\infty f(u) g(u) \,du \,.
\label{eq:inner}
\end{equation}
For this definition, we can find that $\phi_k^{\ell m}$ in 
Eqs.~\eqref{eq:fit-formula-QNM1} 
and~\eqref{eq:fit-formula-QNM2} is normalized as 
$(\phi_k^{\ell m},\,\phi_k^{\ell m})=1$.

Using the inner product in Eq.~\eqref{eq:inner} and 
the standard Gram–Schmidt procedure, 
we obtain the orthonormal set 
of mode functions $\tilde{\phi}_k^{\ell m}$. 
Their explicit listing is 
\begin{eqnarray}
\tilde{\phi}_0^{\ell m} &\equiv& \phi_0^{\ell m} \,, \\
\tilde{\phi}_1^{\ell m} &\equiv& 
\frac{1}{\sqrt{1-(\tilde{\phi}_0^{\ell m},\,\phi_1^{\ell m})^2}} \left[ \phi_1^{\ell m}
- (\tilde{\phi}_0^{\ell m},\,\phi_1^{\ell m})
\,\tilde{\phi}_0^{\ell m} \right] \,, \\
\tilde{\phi}_2^{\ell m} &\equiv& 
\frac{1}{\sqrt{1
-(\tilde{\phi}_1^{\ell m},\,\phi_2^{\ell m})^2
-(\tilde{\phi}_0^{\ell m},\,\phi_1^{\ell m})^2}}
\cr && \times
\left[ \phi_2^{\ell m}
- (\tilde{\phi}_1^{\ell m},\,\phi_2^{\ell m})
\,\tilde{\phi}_1^{\ell m} 
- (\tilde{\phi}_0^{\ell m},\,\phi_2^{\ell m})
\,\tilde{\phi}_0^{\ell m}\right] \,, \\
\tilde{\phi}_3^{\ell m} &\equiv& 
\frac{1}{\sqrt{1
-(\tilde{\phi}_2^{\ell m},\,\phi_3^{\ell m})^2
-(\tilde{\phi}_1^{\ell m},\,\phi_3^{\ell m})^2
-(\tilde{\phi}_0^{\ell m},\,\phi_3^{\ell m})^2}}
\cr && \times
\left[ \phi_3^{\ell m} 
- (\tilde{\phi}_2^{\ell m},\,\phi_3^{\ell m})
\,\tilde{\phi}_2^{\ell m} 
- (\tilde{\phi}_1^{\ell m},\,\phi_3^{\ell m})
\,\tilde{\phi}_1^{\ell m} 
- (\tilde{\phi}_0^{\ell m},\,\phi_3^{\ell m})
\,\tilde{\phi}_0^{\ell m}
\right] \,, \\
\vdots && \cr
\tilde{\phi}_k^{\ell m} &\equiv&
\left[ 
1-\sum_{p=0}^{k-1} (\tilde{\phi}_p^{\ell m}, \phi_k^{\ell m})^2
\right]^{-1/2}
\left[ \phi_k^{\ell m}
- \sum_{p=0}^{k-1} (\tilde{\phi}_p^{\ell m}, \phi_k^{\ell m}) 
\, \tilde{\phi}_p^{\ell m}
\right]\,.
\end{eqnarray}
However, we note that a direct numerical implementation of above construction
does not work well because of the accumulation of round-off errors. 
In practice, we use the modified Gram-Schmidt 
algorithm~\cite{GoluVanl96}
to obtain accurately the orthonormal set of mode functions.

From the above orthonormal set of mode functions $\tilde{\phi}_k^{\ell m}$,
we can extract the fitting coefficients easily by
\begin{equation}
\tilde{C}_k^{\ell m} = (\tilde{\phi}_k^{\ell m},\, \Psi^{\ell m}_{\rm num}) \,,
\label{eq:mode-coeffs}
\end{equation}
from the numerically generated Zerilli-Moncrief function $\Psi^{\ell m}_{\rm num}$,
and construct the modified model waveform as
\begin{equation}
\tilde{\Psi}_N^{\ell m}(u) =
\sum_{k=0}^{2N+1} \tilde{C}_k^{\ell m} \tilde{\phi}_k^{\ell m}(u) \,,
\label{eq:fit-formula-QNM-ON}
\end{equation}
up to a finite overtone index of $n=N$.

\section{Results}
\label{sec:results}

We now present the demonstration of the QNM fits including 
overtones given in Eqs.~\eqref{eq:fit-formula-QNM} 
and~\eqref{eq:fit-formula-QNM-ON} with the close-limit TD waveform 
for head-on collisions explained in Section~\ref{sec:method}.

\subsection{Quasinormal-mode fitting and residuals}

\begin{figure}[ht]
\centering
\includegraphics[width=0.45\textwidth]{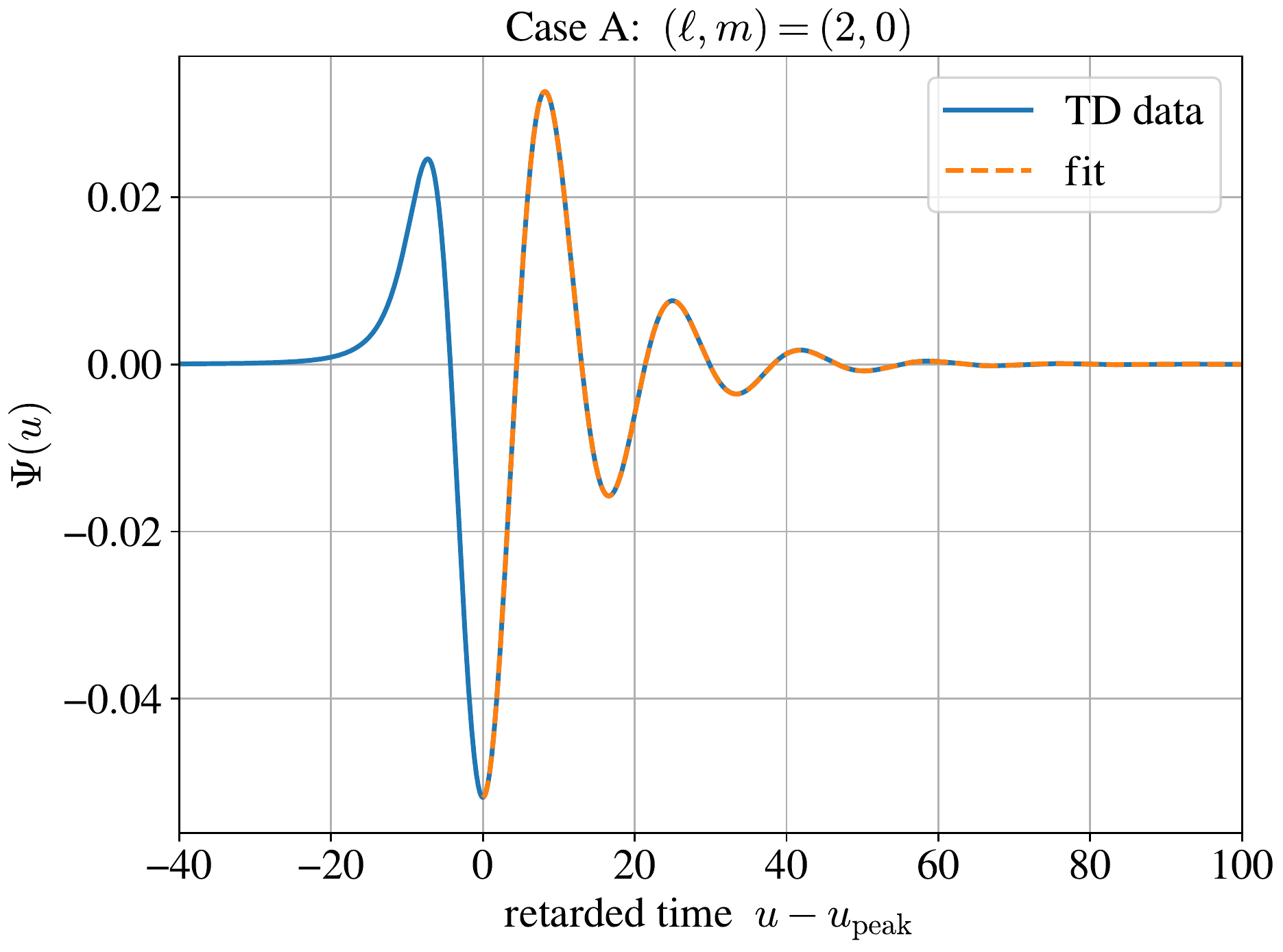} \\
\includegraphics[width=0.45\textwidth]{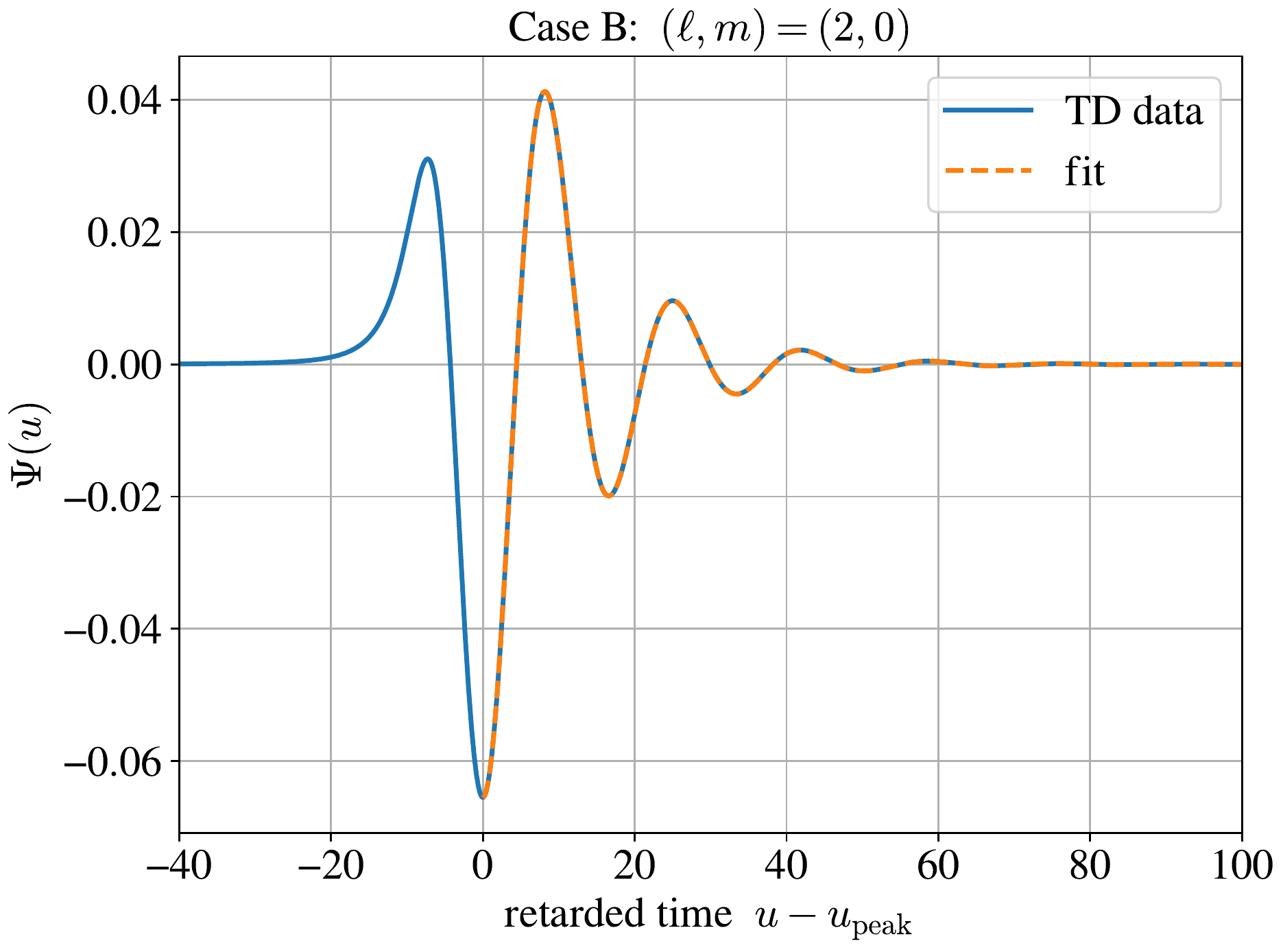}
\includegraphics[width=0.45\textwidth]{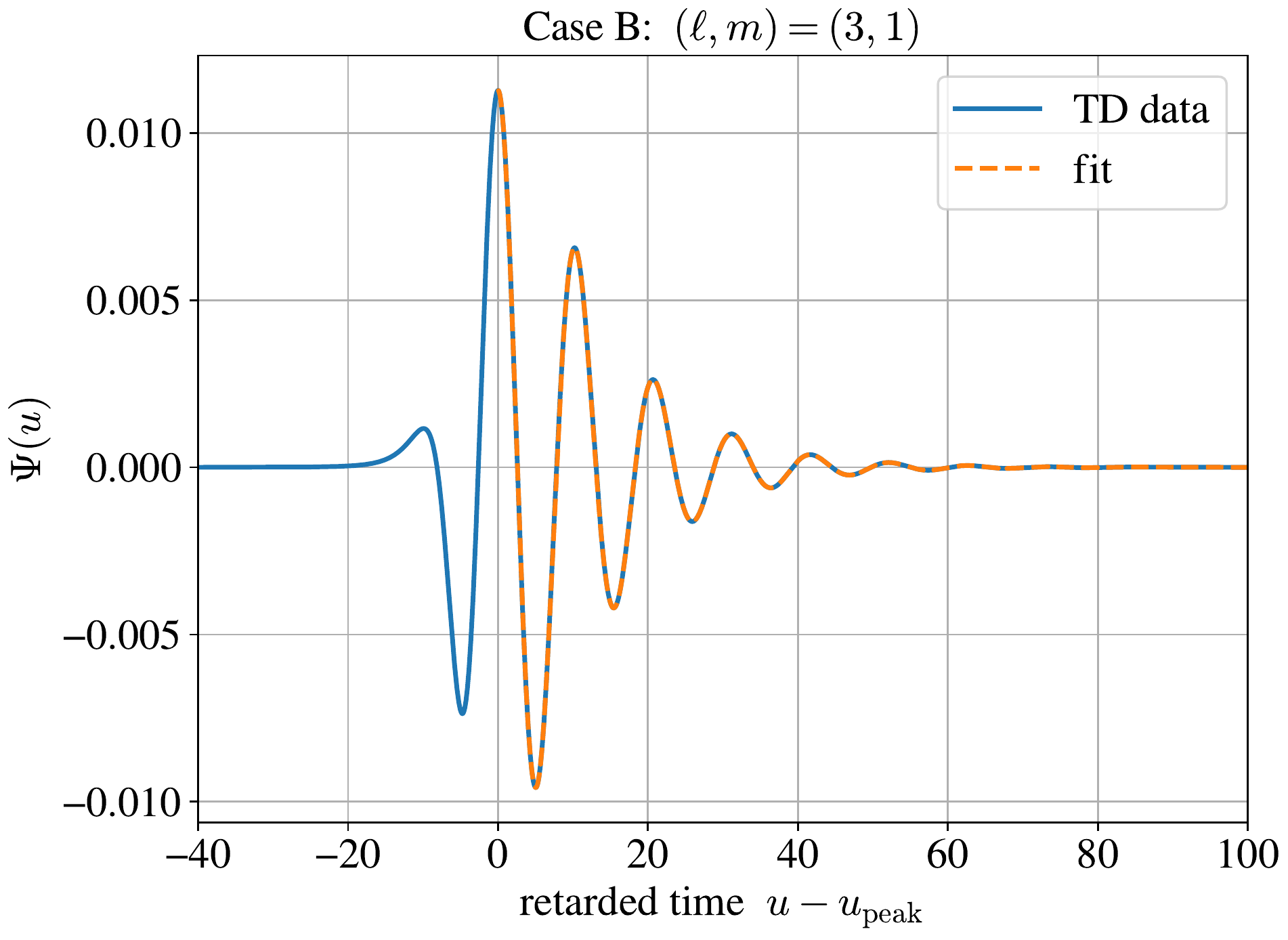}
\caption{TD data of the Zerilli-Moncrief functions $\Psi^{\ell m}$ 
with the 2PN close-limit initial data for the head-on collision 
(the blue solid lines) and QNM fitting with $N=7$ 
in Eq.~\eqref{eq:fit-formula-QNM-ON} 
(the orange dashed lines).
The TD data is extracted as a function of
the retarded time, $u$ on the $v=v_f=3000M$ surface.
The top panel shows the $(\ell=2,\,m=0)$ mode computed for the case A.
The bottom two panel shows 
the $(\ell=2,\,m=0)$ mode (left) and
the $(\ell=3,\,m=1)$ mode (right) computed for the case B.}
\label{fig:headon-waveform}
\end{figure}

Figure~\ref{fig:headon-waveform} shows the QNM fit
including overtones in Eq.~\eqref{eq:fit-formula-QNM-ON} to the $(\ell=2,\,m=0)$ mode 
of the Zerilli-Moncrief function $\Psi^{\ell m}$ 
in the case A (top). 
Similarly, in the case B, 
we show the QNM fits to the $(\ell=2,\,m=0)$ mode (bottom-left) 
and to the $(\ell=3,\,m=1)$ mode (bottom-right),
respectively. 
In each panel, the QNM overtone in the model waveform is truncated at $N=7$
and the time of the peak amplitude, $u_\textrm{peak}$ 
is chosen as the starting time, $u_0$.

\begin{figure}[ht]
\centering
\includegraphics[width=0.45\textwidth]{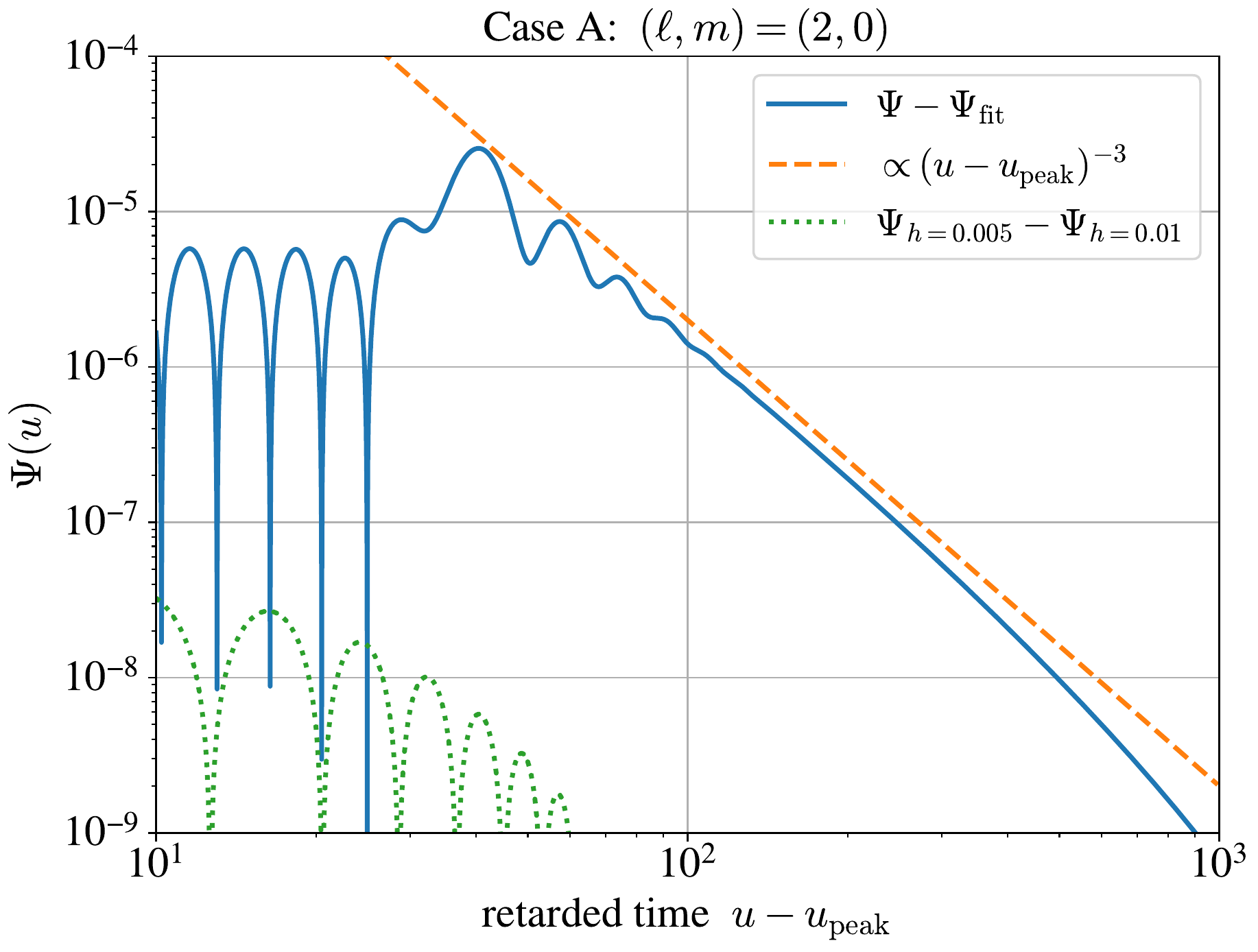} \\
\includegraphics[width=0.45\textwidth]{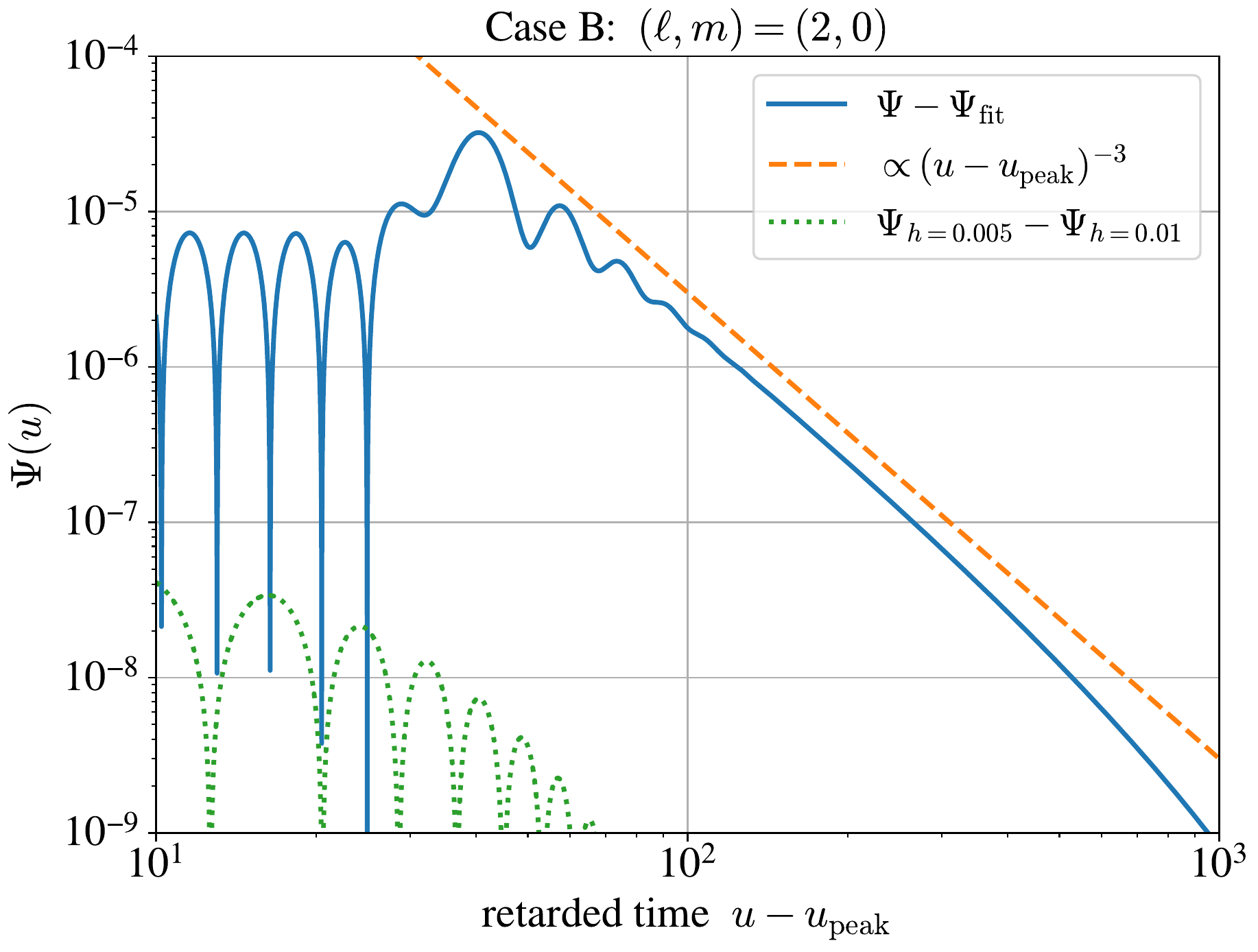}
\includegraphics[width=0.45\textwidth]{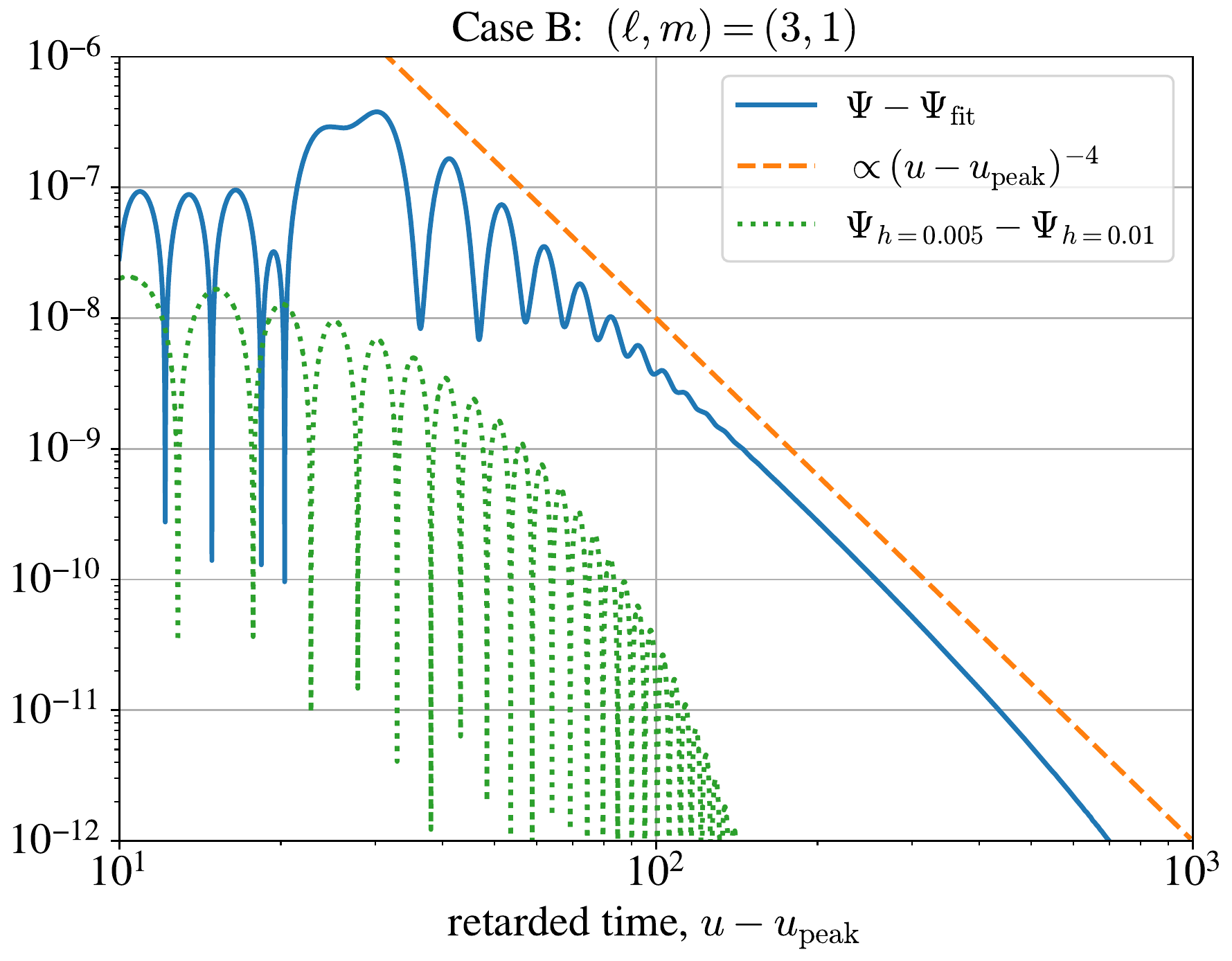}
\caption{Fit residuals (the blue solid lines) 
between the Zerilli-Moncrief TD data 
$\Psi^{\ell m}_{\rm num}$ and 
the QNM fit including overtones with $N = 7$. 
The grouping of panels are the same as for Figure~\ref{fig:headon-waveform}.
For reference, we plot $\Delta\Psi$ in
Eq.~\eqref{eq:TDerror-estimator}
with $h_1=0.01M$ and $h_2=0.005M$ (the green dotted lines) 
to roughly estimates the numerical error in the data. 
We also show the lines proportional to
$(u-u_\textrm{peak})^{-3}$ in the top and bottom-left panel and to $(u-u_\textrm{peak})^{-4}$ in the bottom-right panel in order to see the
late-time behavior (the orange dashed lines).}
\label{fig:headon-waveform_2}
\end{figure}

Figure~\ref{fig:headon-waveform_2} provides the QNM fit residuals of 
Figure~\ref{fig:headon-waveform} (the blue solid lines). 
We find that the residuals are all larger than our numerical uncertainty 
(the green dotted lines), and shows the power-law behavior 
in the late time ($u-u_\textrm{peak}>100M$, the orange dashed lines). 
This corresponds to the late-time tail in the evolution of 
a perturbation of the gravitational field 
around a remnant Schwarzschild BH, 
studied in various literature; 
see, e.g., Refs.~\cite{Leaver:1986gd,Ching:1995tj,Andersson:1996cm} 
for the Schwarzschild background spacetime, 
and, e.g., Refs.~\cite{Krivan:1997hc,Pazos-Avalos:2004uyd} 
for the Kerr background spacetime. 
The power-law tail in a sufficiently late time is also investigated  
using a sophisticated numerical method
with compactification along hyperboloidal surfaces~\cite{Zenginoglu:2008uc}.
Each of QNM fit residuals deviates from the power-law line 
when $u$ becomes larger because the null infinity approximation $v \gg u$ is broken
(see Section~IV in Ref.~\cite{Gundlach:1993tp}).

\begin{figure}[ht]
\centering
\includegraphics[width=0.45\textwidth]{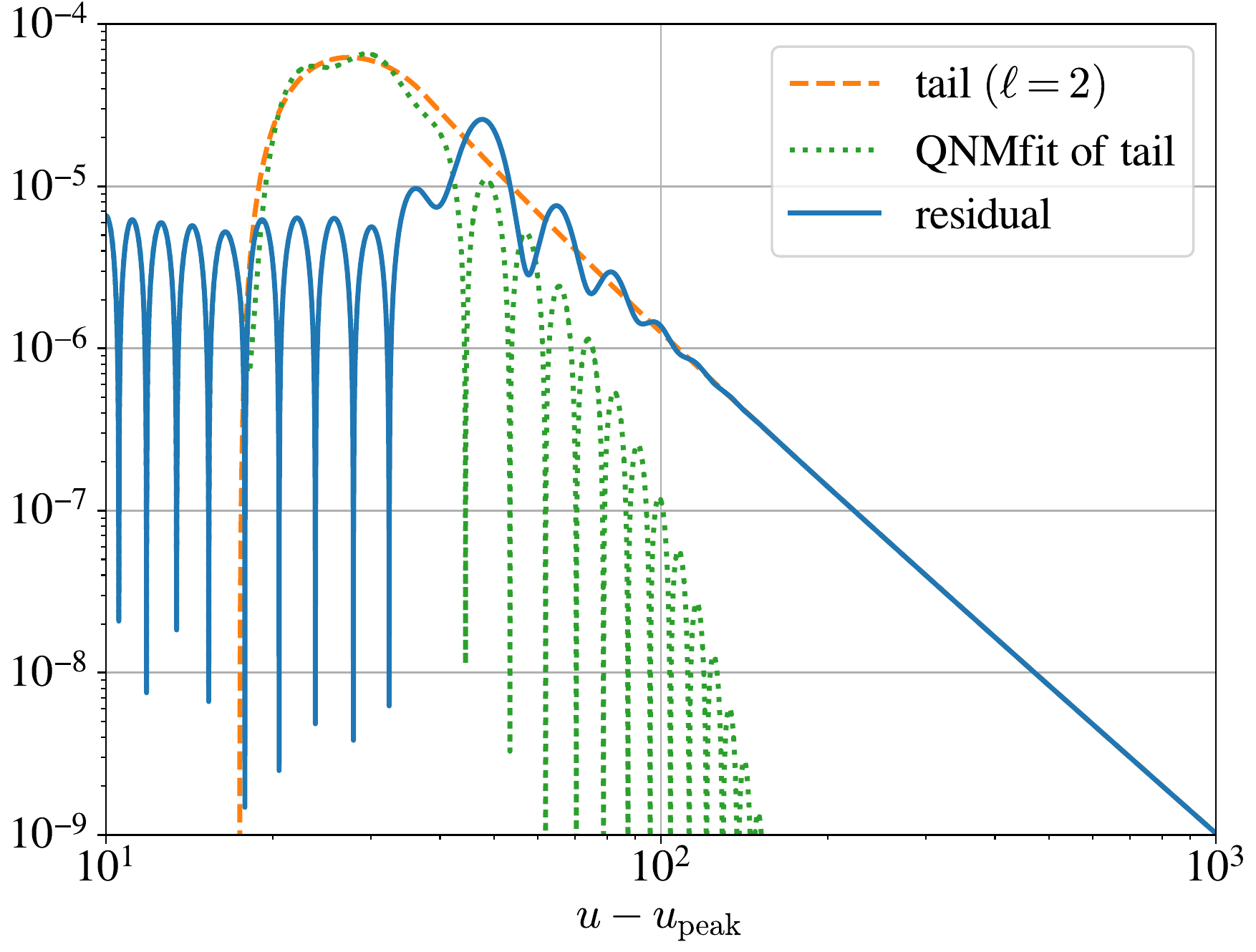}
\includegraphics[width=0.45\textwidth]{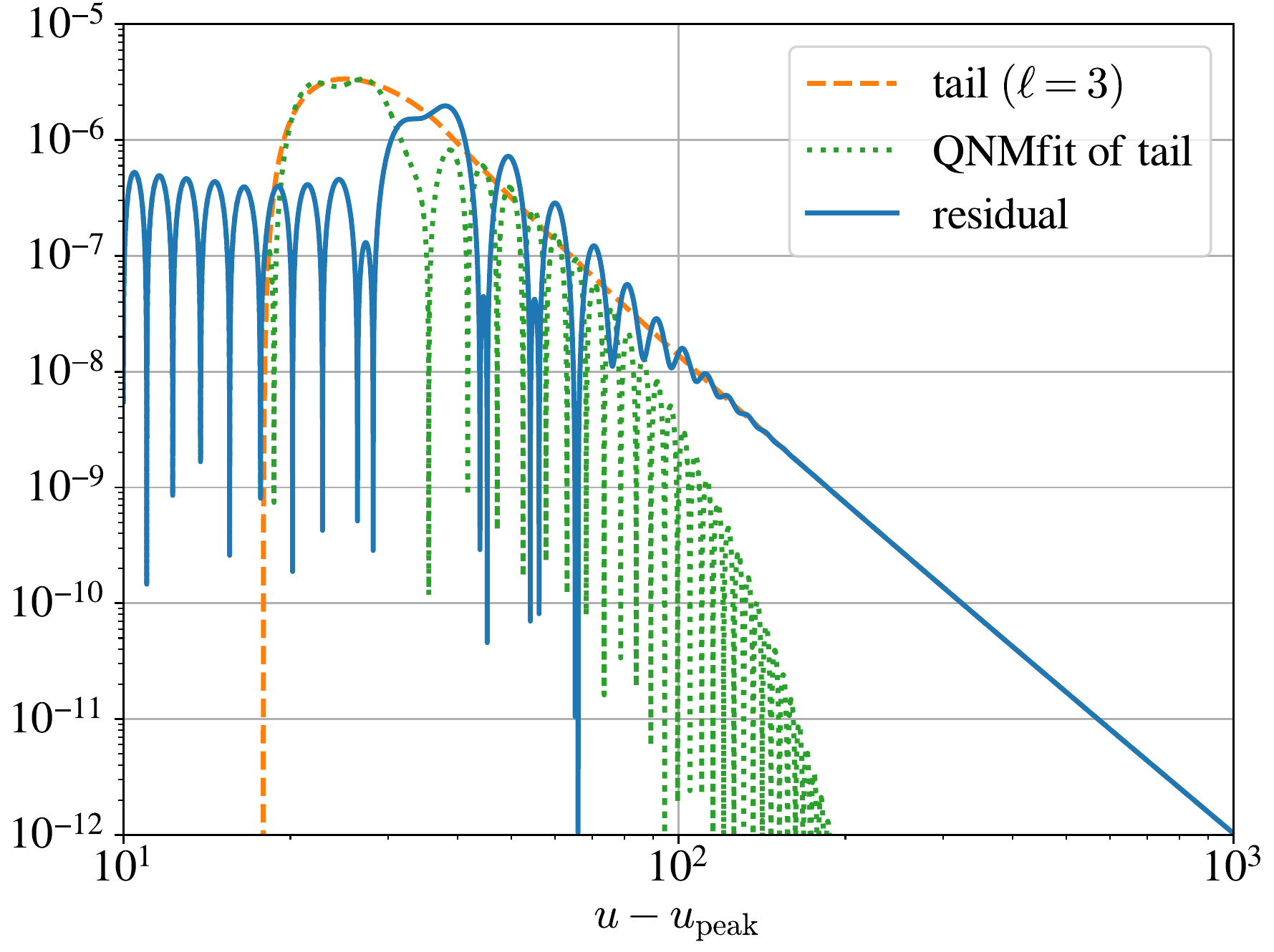}
\caption{Analysis of mock data of the tail component
in Eq.~\eqref{eq:pseudo_tail} (the orange dashed lines). 
We plot the $(\ell=2)$ and $(\ell=3)$ modes 
in the left and right panels, respectively.
The green dotted lines show the QNM fit including overtones with $N = 7$
and the blue solid lines are the fit residuals. 
Here, we choose $D_l=1$, $u_1=u_\textrm{peak}+10M$ and
$u_2=u_\textrm{peak}+30M$
in Eq.~\eqref{eq:pseudo_tail} with Eq.~\eqref{eq:WF}.}
\label{fig:Qfit_to_tail}
\end{figure}

In Figure~\ref{fig:headon-waveform_2}, 
we also find that the QNM fit residuals in the early-time
($u-u_\textrm{peak} \lesssim 100M$) 
do not improve even if one includes more overtones 
into the model waveform in 
Eq.~\eqref{eq:fit-formula-QNM-ON}. 
Although we do not have a mathematically rigorous proof, 
we speculate that this limitation may arise from 
the QNM fit including overtones
(that describes the exponential decay of the perturbation) 
to the power-law tail component of the perturbation, 
based on a simple numerical experiment detailed below.

First, we prepare a mock data of the tail component of the perturbation as
\begin{equation}
\Psi^{\ell m}_\textrm{tail}(u) =
D_\ell W(u)
\left(\frac{u-u_\textrm{peak}}{M}\right)^{-\ell-1} \,,
\label{eq:pseudo_tail}
\end{equation}
where $D_{\ell}$ is an arbitrary constant,
and we define a window function by
\begin{equation}
W(u) = \left\{ \begin{array}{ll}
0 & (u<u_1) \,, \\
\dfrac{(u-u_1)^2(2u+u_1-3u_2)}{(u_1-u_2)^3} 
& (u_1\le u \le u_2)  \,, \\
1 & (u_2<u) \,,
\end{array} \right.
\label{eq:WF}
\end{equation}
with $u_\textrm{peak}<u_1<u_2$. 
The window function is introduced to remove the divergent behavior of 
the late-time tail at $u=u_\textrm{peak}$.

Next, we fit this mock data using the QNM model 
including overtones in Eq.~\eqref{eq:fit-formula-QNM-ON}
in the same manner as we do for the TD data. 
Figure~\ref{fig:Qfit_to_tail} shows the QNM fit 
including overtones
to the mock tail data and corresponding fit residuals. 
We see that the residuals in the early-time ($(u-u_\textrm{peak})/M \lesssim 100$) 
show the similar behaviour in Figure~\ref{fig:headon-waveform_2}. 
This result supports our expectation 
about the early-time residual in the TD data.

\subsection{Convergence of the fitting coefficients}

\begin{figure}[ht]
\centering
\includegraphics[width=0.45\textwidth]{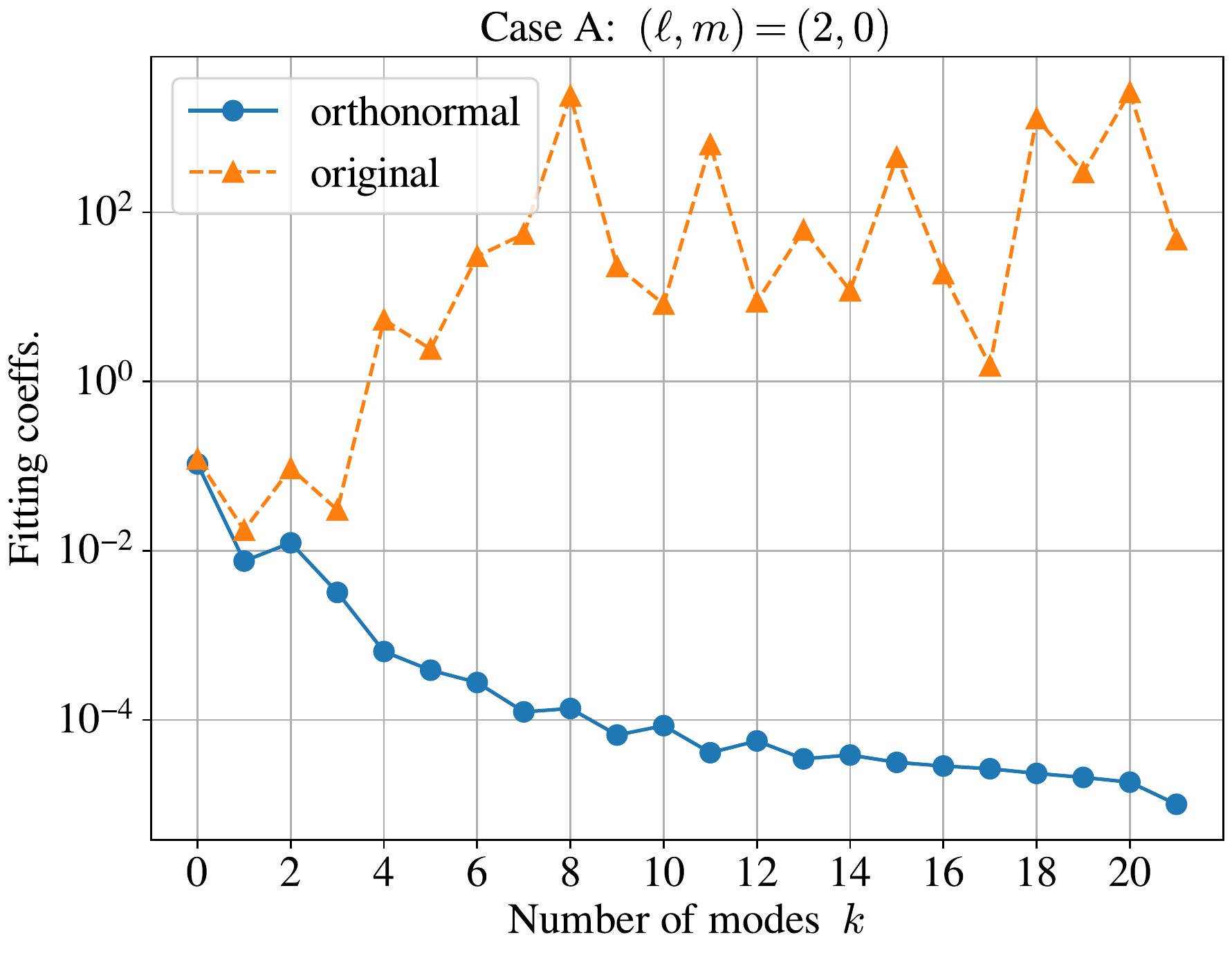}
\includegraphics[width=0.45\textwidth]{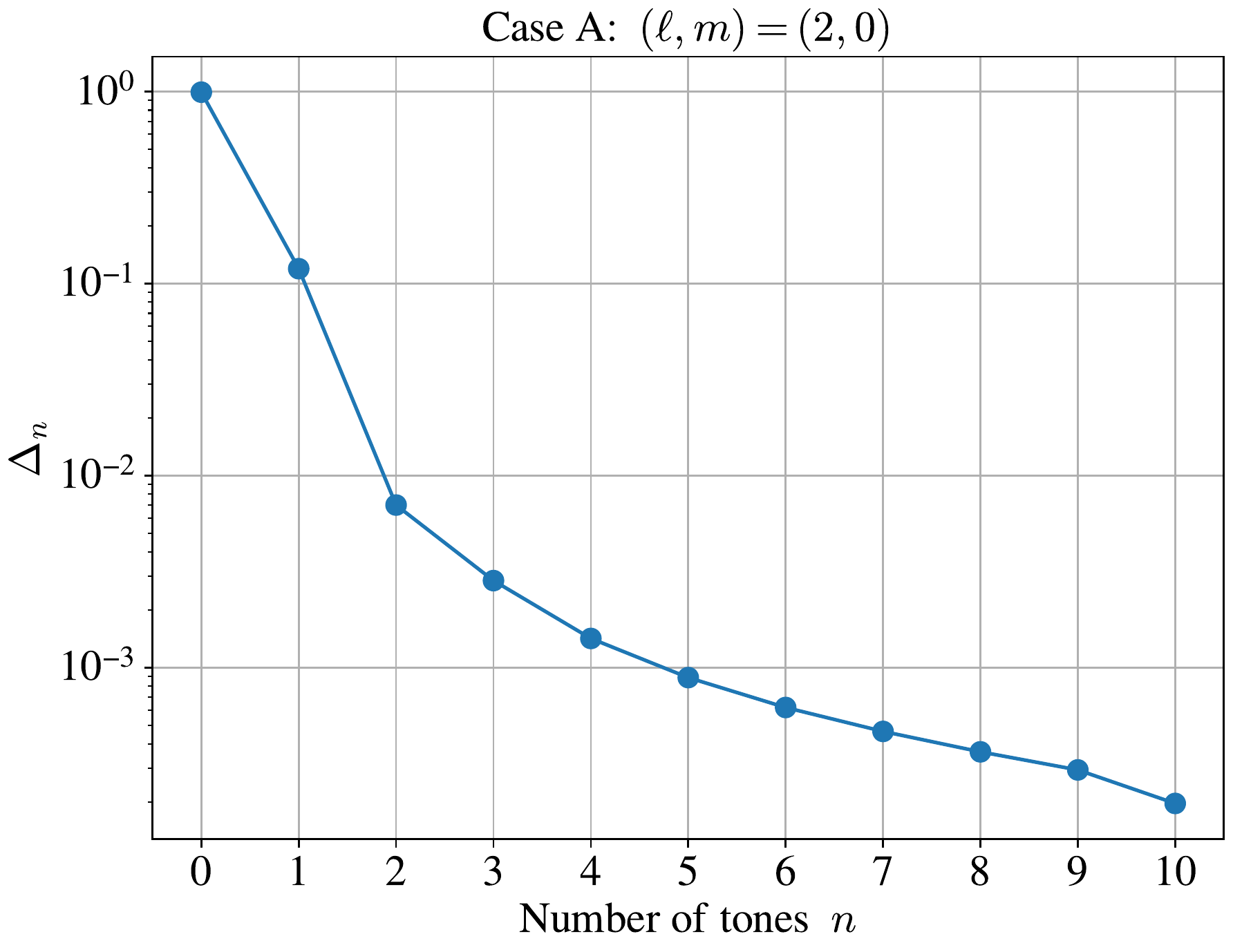}
\caption{\textit{Left}: 
the absolute values of the fitting coefficients, $\tilde{C}_k^{\ell m}$ 
(the blue filled circles) and $C_k^{\ell m}$ (the orange filled triangles) 
for the $(\ell=2,\,m=0)$ mode in the case A, 
as a function of mode $k$.
$\tilde{C}_k^{\ell m}$ is calculated by Eq.~\eqref{eq:mode-coeffs}, 
while $C_k^{\ell m}$ is obtained by the least square fit of the TD
data to the model in Eq.~\eqref{eq:fit-formula-QNM}.
\textit{Right}: the estimator $\Delta_n$ in Eq.~\eqref{eq:Delta_n} 
as a function of overtones $n$.}
\label{fig:coeffs-convergence}
\end{figure}

Because the model waveform 
in Eqs.~\eqref{eq:fit-formula-QNM} 
and~\eqref{eq:fit-formula-QNM-ON} involves 
many overtones $k$ (or $N$)
it is instructive to check their convergence 
in terms of fitting coefficients on $k$, 
to ask how higher overtones should be included in practice 
when we wish to model the TD data from the peak time at 
$u = u_\textrm{peak}$.

In the left panel of Figure~\ref{fig:coeffs-convergence}
where the TD data is for the $(\ell=2,\,m=0)$ mode 
in the case A,
we show the fitting coefficients for the original QNM basis 
$C_k^{\ell m}$ (the orange filled triangles)
and those for the orthonormal set of mode functions
$\tilde{C}_k^{\ell m}$ (the blue filled circles)
as a function of overtones up to $k\le 21$ (i.e., $N=10$). 
It is found that $|\tilde{C}_k^{\ell m}|$ decreases 
roughly in power law of $k$, 
while $|C_k^{\ell m}|$ does not. 

Due to the benefit of the orthonormalization of mode functions
in Eq.~\eqref{eq:fit-formula-QNM-ON}, 
it can be conveniently used to assess the fraction of
power in each mode
\begin{equation}
\rho^2 \equiv (\tilde{\Psi}_N^{\ell m}, \tilde{\Psi}_N^{\ell m})
= \sum_{k=0}^{2N+1} |\tilde{C}_k^{\ell m}|^2 \,.
\end{equation}
Based on the fact that one more mode is added to the orthonormal set 
when $k$ increases by two, we introduce an estimator 
to characterize the match (or overlap) between the TD data 
and the QNM fit model including overtones at a given $n$
\begin{equation}
\Delta_n \equiv 
\frac{\sqrt{|\tilde{C}_{2n}^{\ell m}|^2
      + |\tilde{C}_{2n+1}^{\ell m}|^2}}{\rho} \,.
\label{eq:Delta_n}
\end{equation}

\begin{figure}[ht]
\centering
\includegraphics[width=0.45\textwidth]{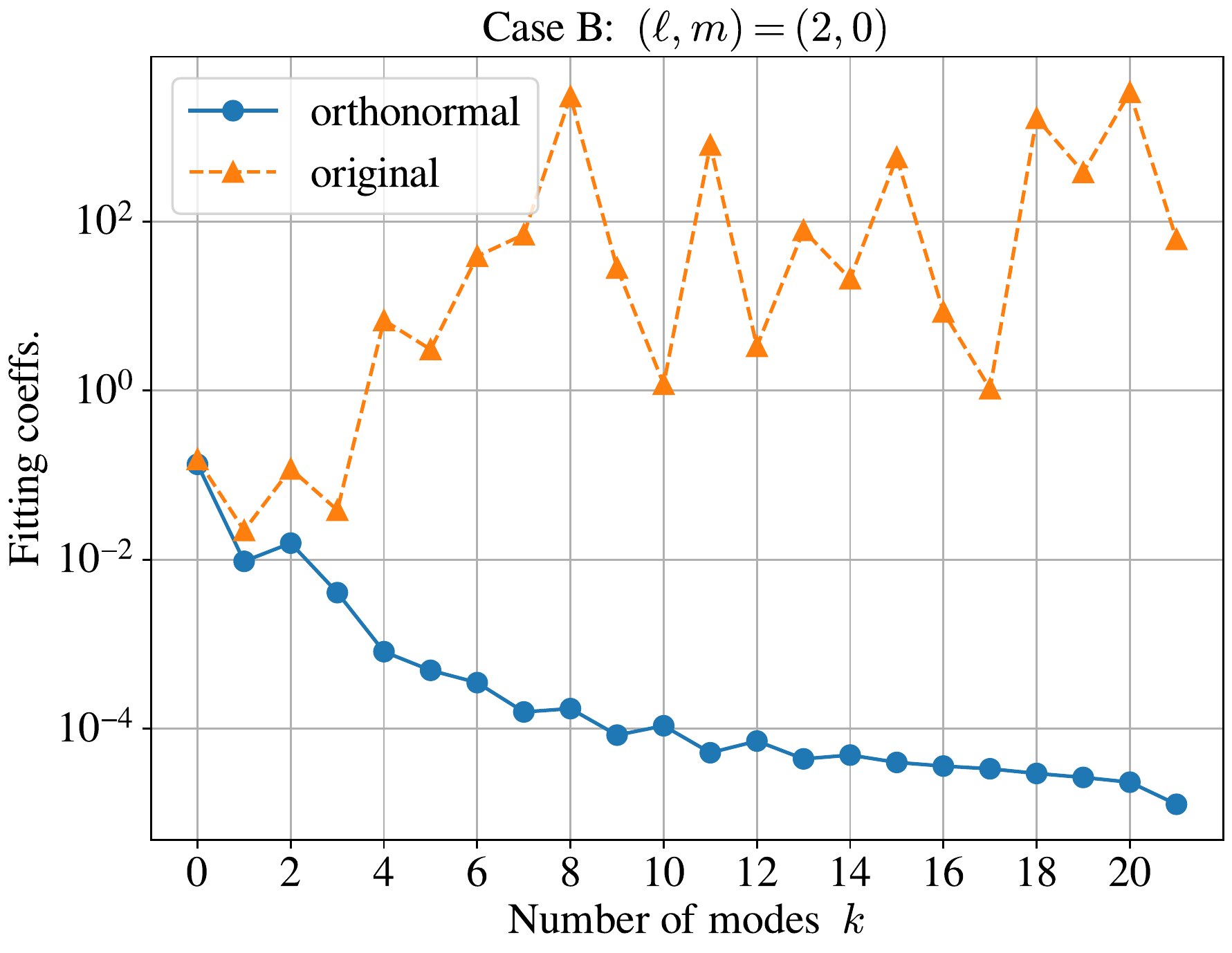}
\includegraphics[width=0.45\textwidth]{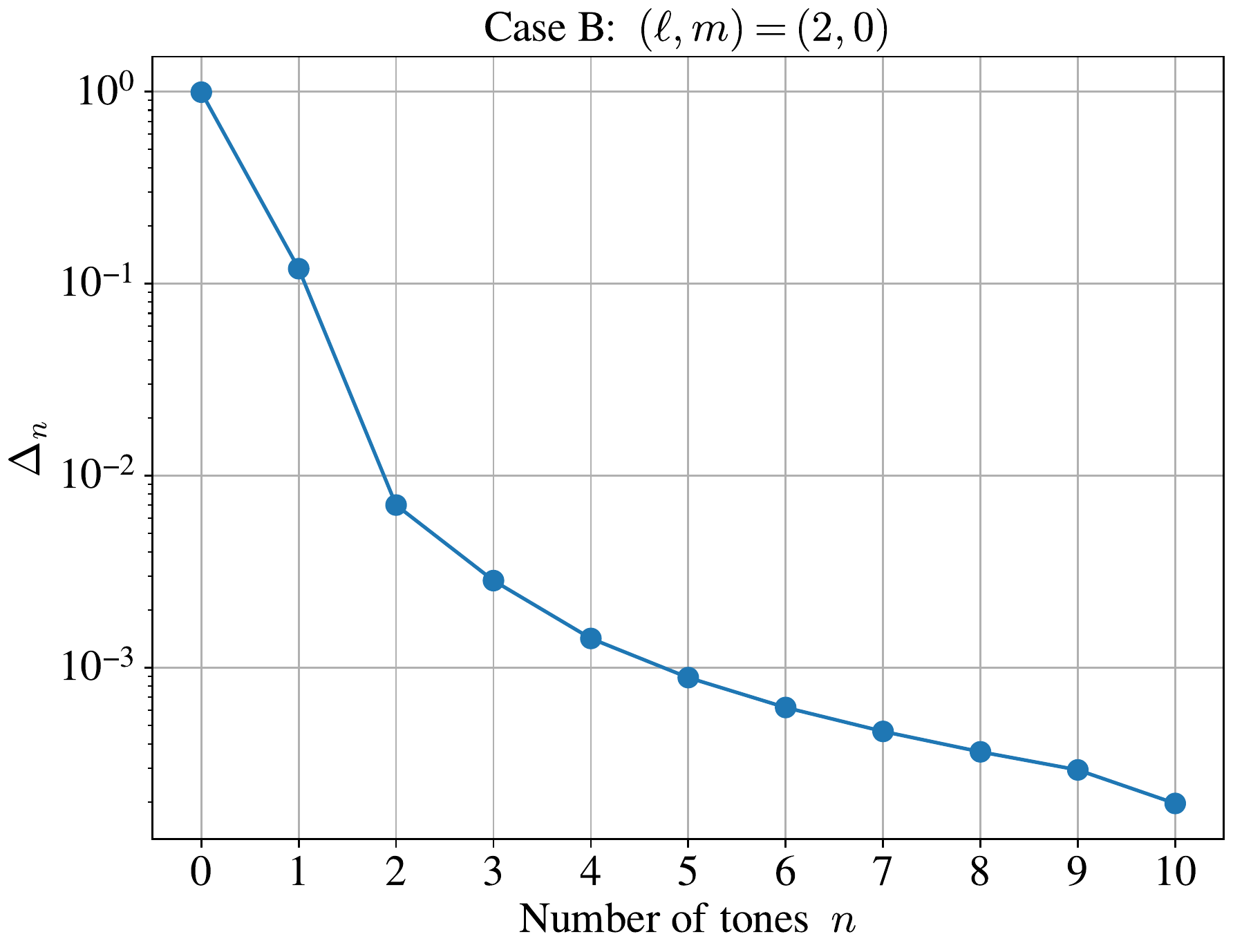} \\
\includegraphics[width=0.45\textwidth]{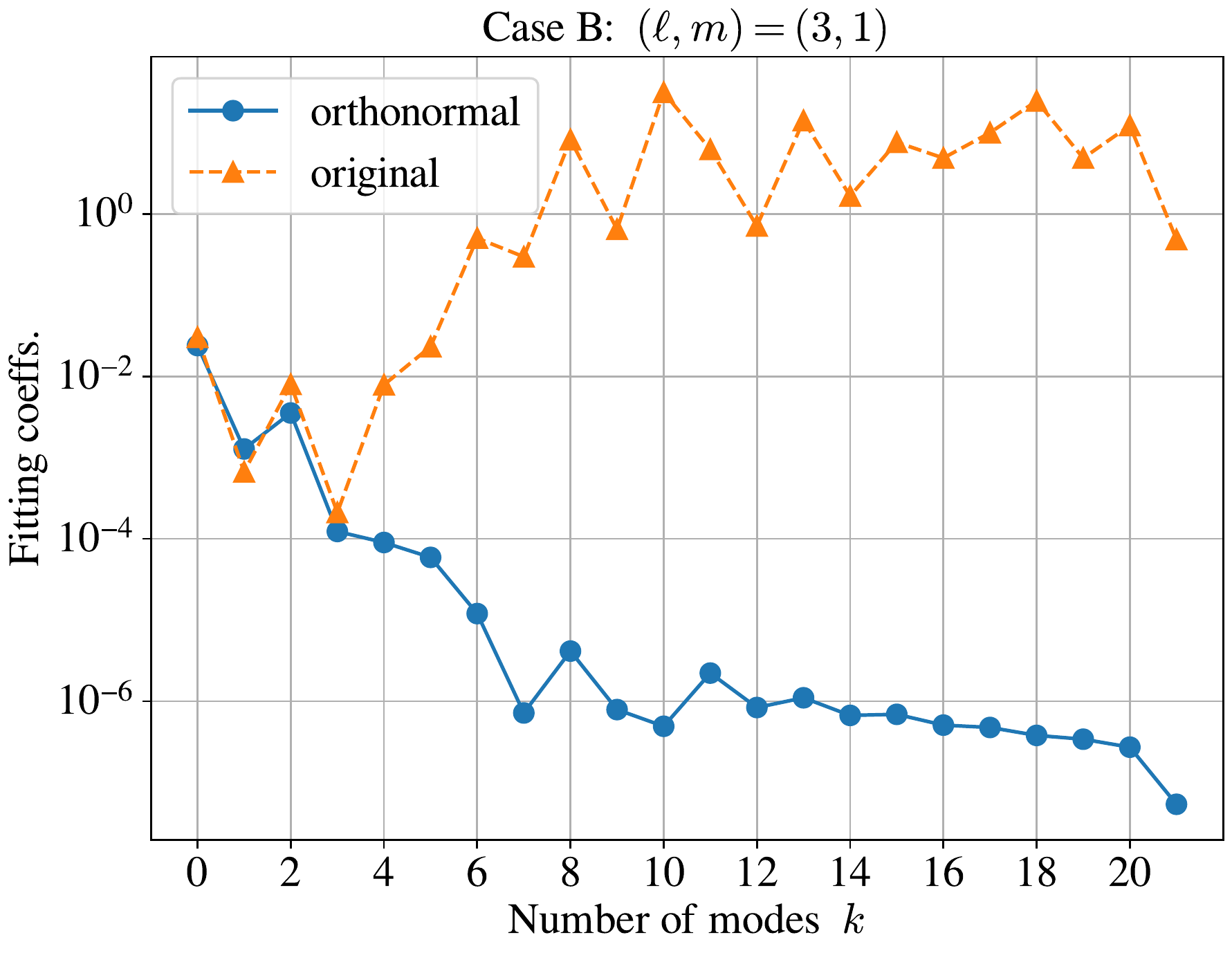}
\includegraphics[width=0.45\textwidth]{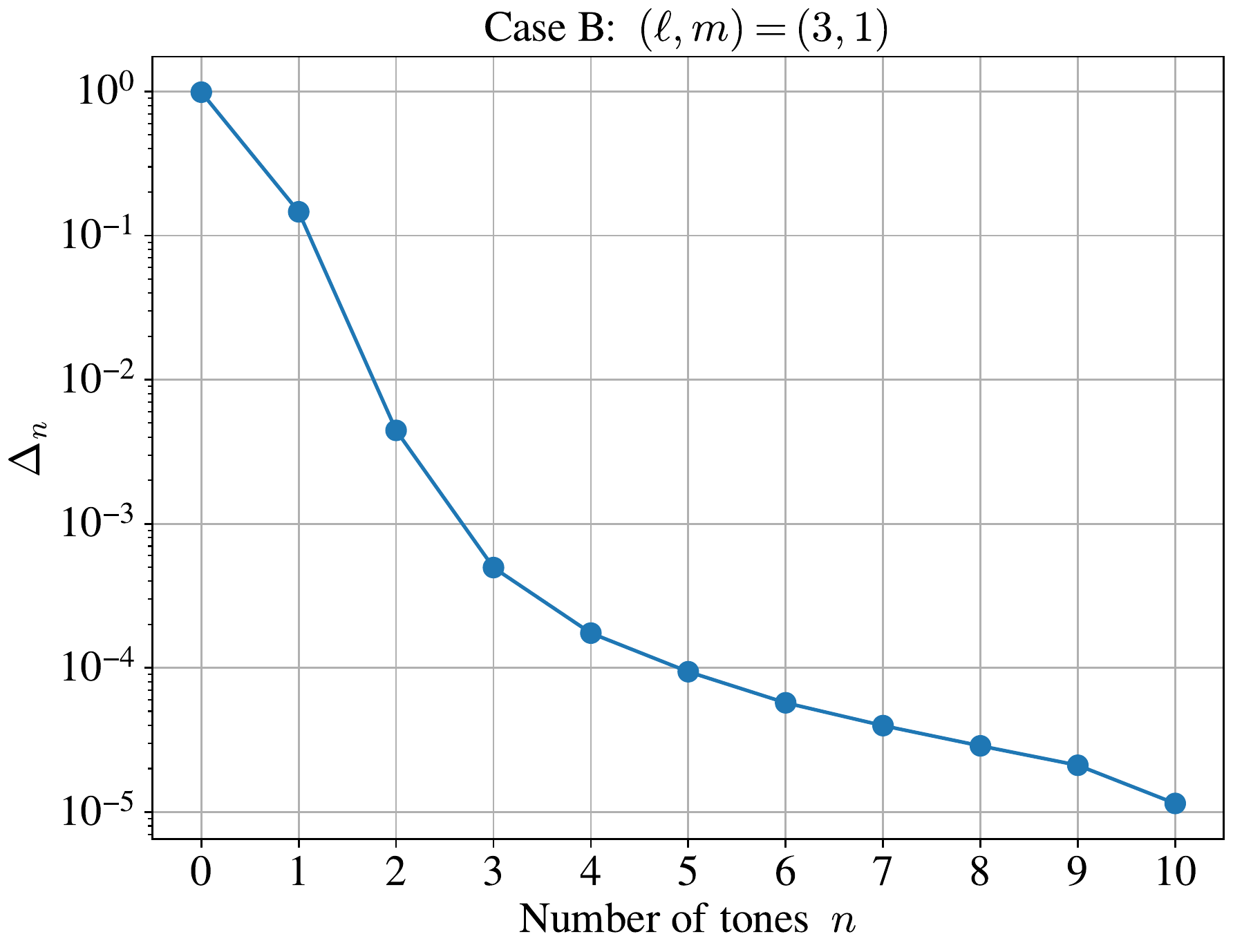}
\caption{The same figure as for Figure~\ref{fig:coeffs-convergence}, 
but using the $(\ell=2,\,m=0)$ mode (top)
and $(\ell=3,\,m=1)$ mode (bottom) in the case B.}
\label{fig:coeffs-convergenceB}
\end{figure}

The right panel of Figure~\ref{fig:coeffs-convergence} shows 
$\Delta_n$ as a function of mode $n$. 
We see that the inclusion of the first overtone ($n=1$) can improve 
the match by $10\%$, and that the QNM fits up to $n=5$ 
will suppress the mismatch less than $0.1\%$.

In the top and bottom panels of Figure~\ref{fig:coeffs-convergenceB}, 
we show the same figures as for Figure~\ref{fig:coeffs-convergence}
but with the $(\ell=2,\,m=0)$ and $(\ell=3,\,m=1)$ modes
in the case B, respectively. 
We see the similar convergent behavior 
in $\tilde{C}_k^{\ell m}$ here, 
and again find that $(n \geq 5)$ higher overtones 
contributes 
to the match only at the level of less than $0.1\%$. 
These results highlight the potential benefit 
to use the orthonormal set of mode functions
in the data analysis of ringdown GWs.

\section{Summary and Discussion}
\label{sec:discussion}

It is now widely recognized the importance to have a ringdown waveform model 
including higher overtones of QNMs (see, e.g., Ref.~\cite{Giesler:2019uxc}). 
With the overtones, one can set the starting time 
of the ringdown GWs
at the time of the peak amplitude much earlier than a time 
($\sim 10M$ -- $20M$ after the peak time)
to obtain unbiased remnant BH parameters
(see, e.g., Figure~5 for GW150914
in Ref.~\cite{LIGOScientific:2016lio}). 
This allows the increased SNR of observed ringdown signals
as well as better parameter estimation of the remnant BH. 

In this paper, we examined in detail the ringdown GWs 
of binary BHs with QNM fits including overtones 
in Eqs.~\eqref{eq:fit-formula-QNM} 
and~\eqref{eq:fit-formula-QNM-ON}, 
using the accurate close-limit waveform 
in the BH perturbation theory as a test bed. 
Our analysis is restricted to the narrow case of head-on collisions 
of two nonspinning BHs 
(based on the $2$PN initial data 
and linear BH perturbation theory), 
but it suffices to highlight some of the key features 
of the full problem, and we found three main results: 
(i) we reconfirm the importance of QNMs overtones 
to fit the ringdown waveforms
after the time of the peak amplitude. 
This agrees with the previous findings 
in literature~\cite{Giesler:2019uxc,Mourier:2020mwa};
(ii) the small contributions of late-time tail exist 
in the fit residuals at the level of $O(10^{-5})$ or below 
(see Figure~\ref{fig:headon-waveform_2}); 
(iii) the fitting coefficients decay with overtones 
when one uses the orthonormal set of mode functions 
in Eq.~\eqref{eq:mode-coeffs} 
(see Figures~\ref{fig:coeffs-convergence}
and~\ref{fig:coeffs-convergenceB}).

\begin{figure}[ht]
\centering
\includegraphics[width=0.45\textwidth]{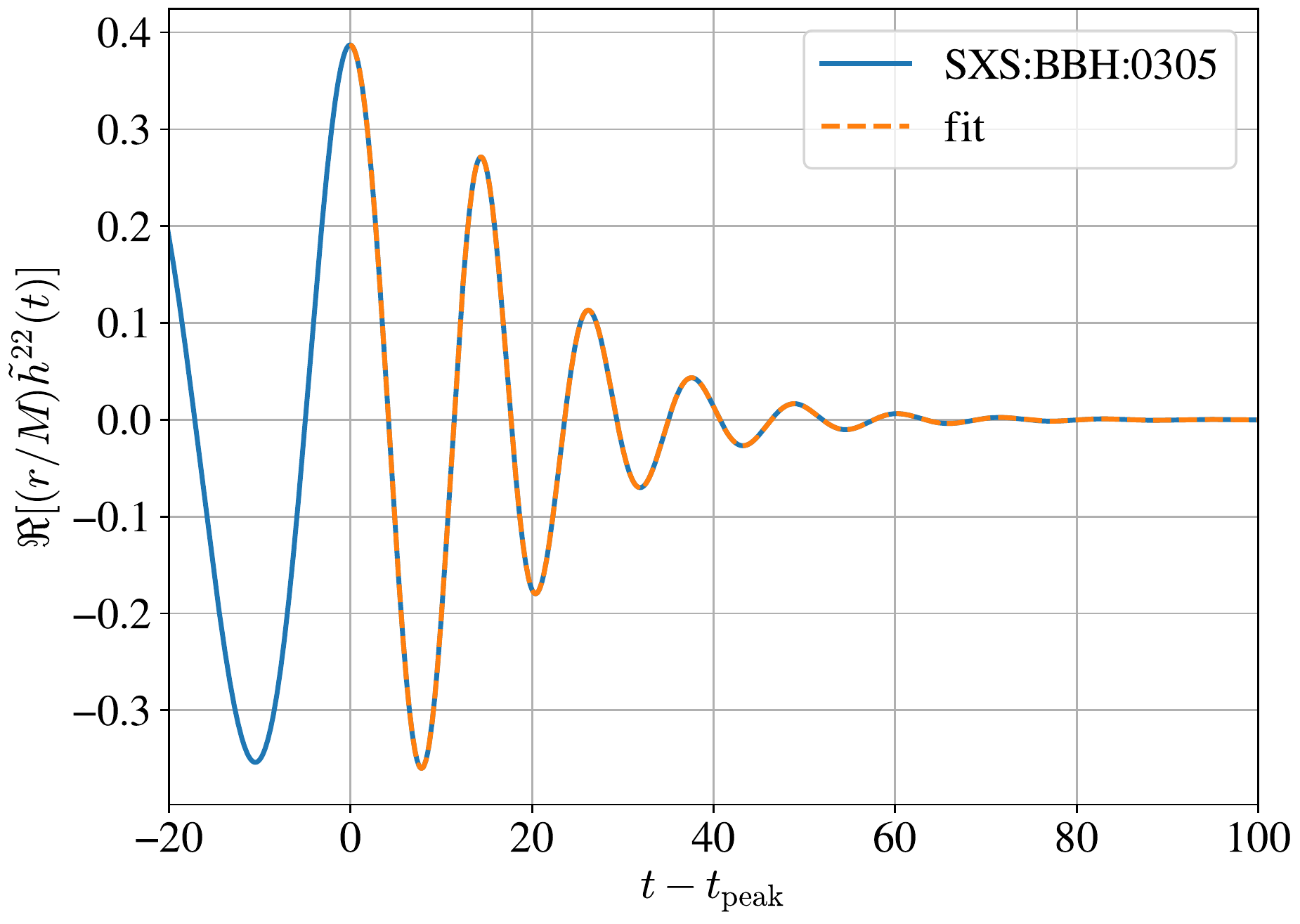}
\includegraphics[width=0.42\textwidth]{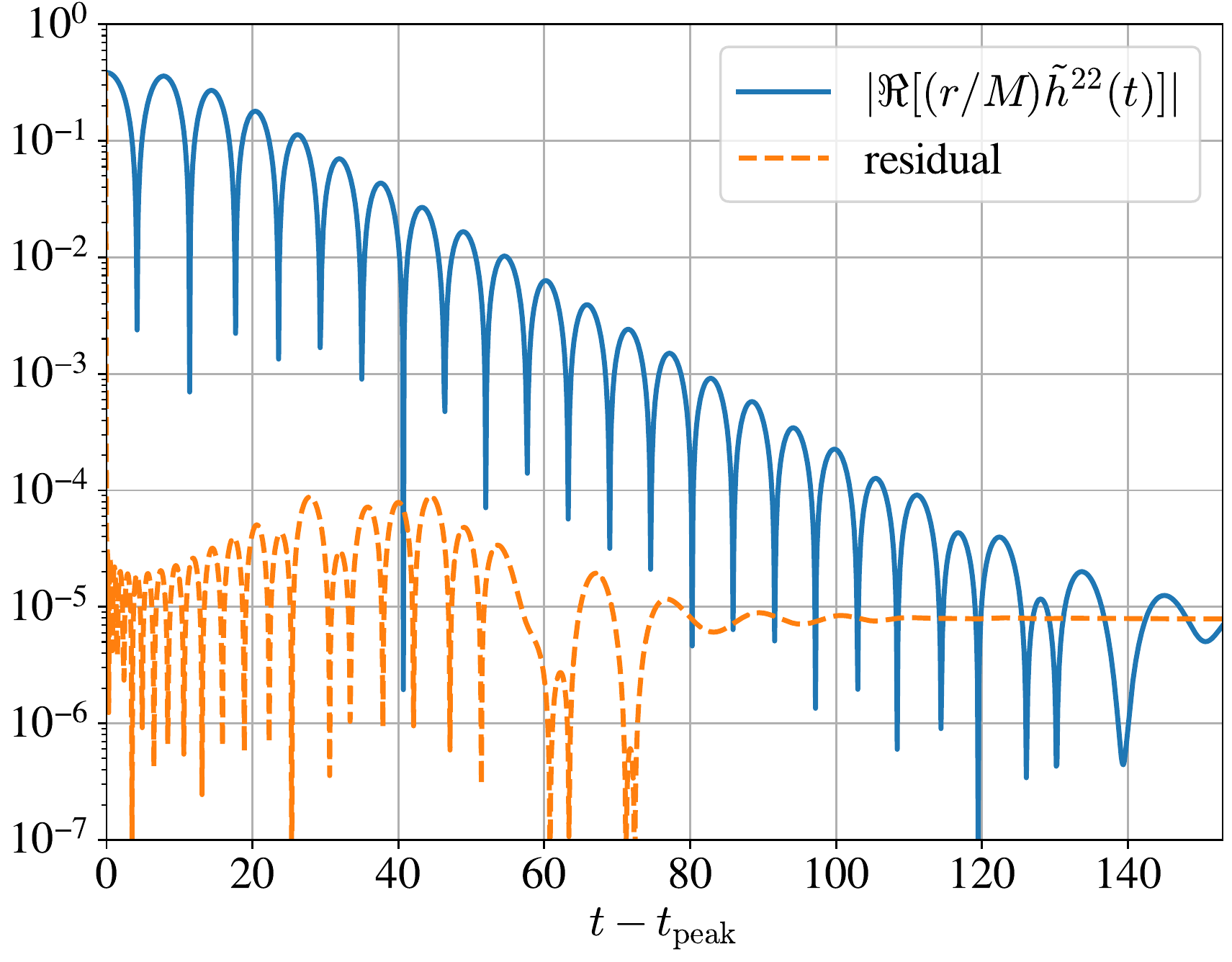}
\caption{\textit{Left}: the $(\ell=2,\,m=2)$ spheroidal harmonic mode
of the NR waveform SXS:BBH:0305 (the blue solid line) 
and the QNM fit including overtone with orthonormalized mode functions 
introduced in Eq.~\eqref{eq:fit-formula-QNM-ON} (the orange dashed line).
\textit{Right}: the NR data (the blue solid line) and
the fit residuals (the orange dashed line).}
\label{fig:BBH0305-waveform}
\end{figure}

\begin{figure}[ht]
\centering
\includegraphics[width=0.45\textwidth]{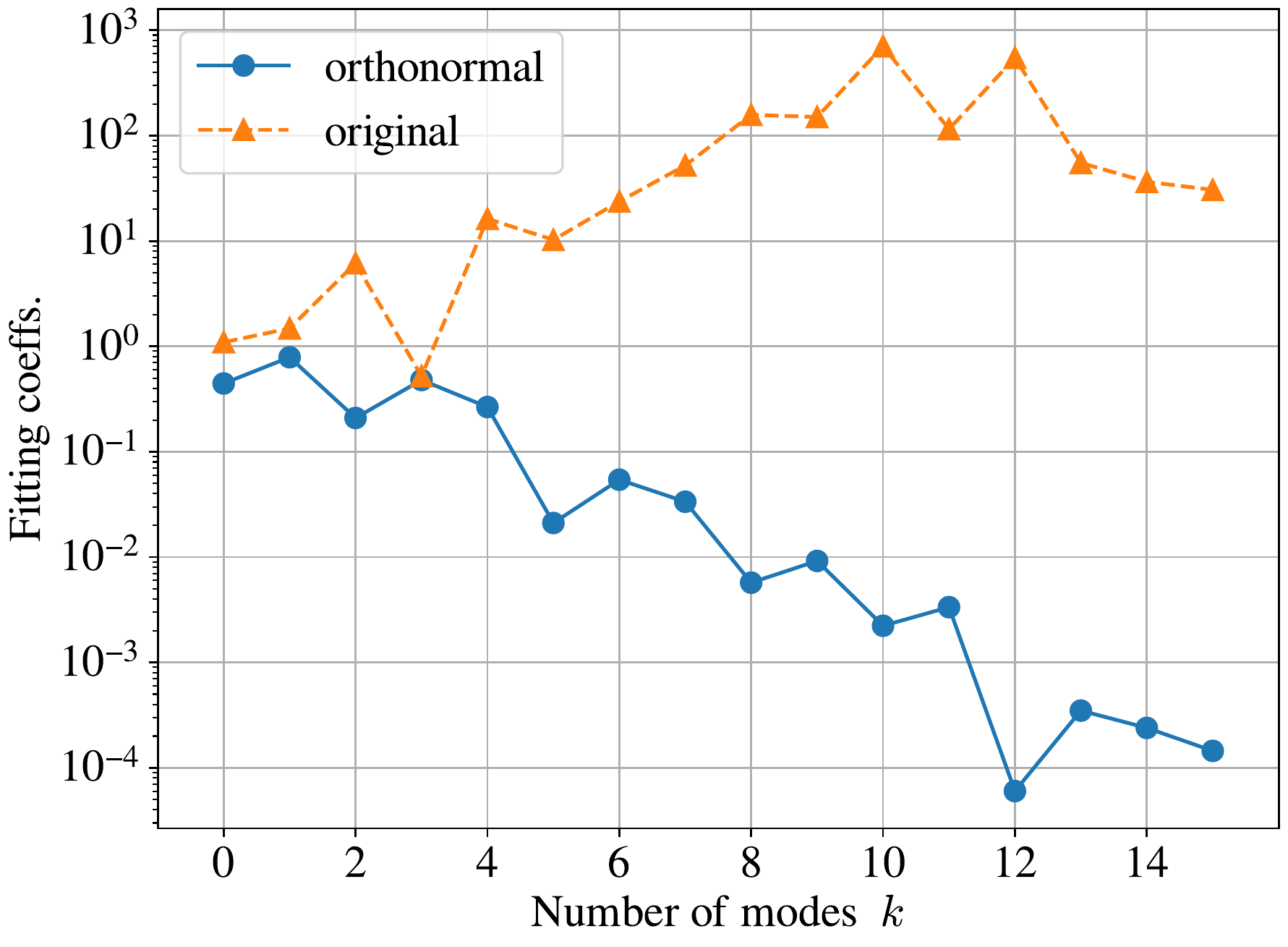}
\includegraphics[width=0.45\textwidth]{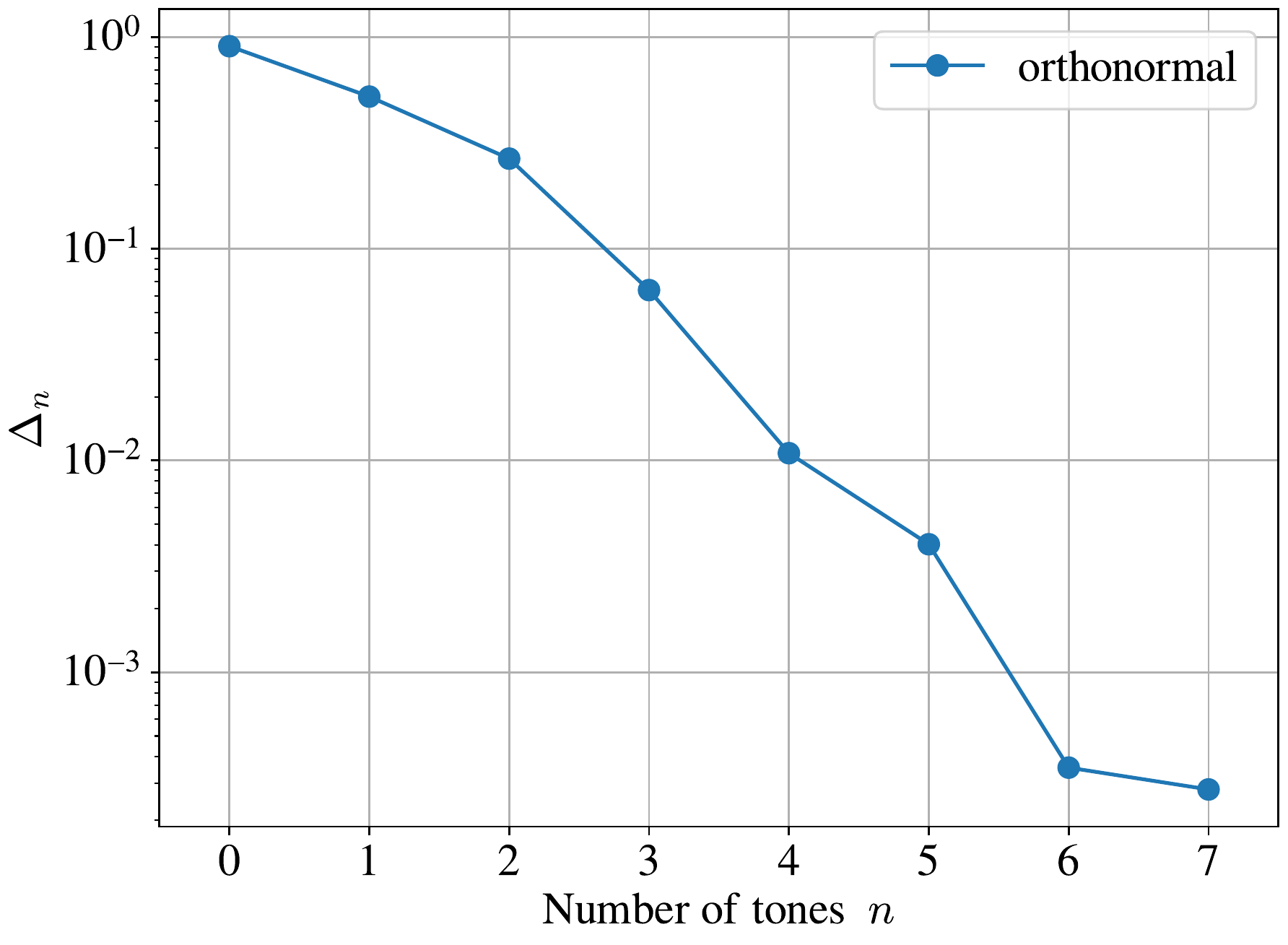}
\caption{ 
Fitting coefficients of the $(\ell=2,\,m=2)$ 
spheroidal harmonic mode of SXS:BBH:0305.
The left panel displays the absolute values of the
fitting coefficients for $\tilde{C}_k^{\ell m}$ 
(the blue filled circles) and $C_k^{\ell m}$ (the orange filled triangles).
The plots in the right column shows $\Delta_n$ in Eq.~\eqref{eq:Delta_n}.}
\label{fig:coeffs-convergence-NR}
\end{figure}

A natural next step of our study is to assess the merit of 
the orthonormal set of mode functions 
in the context of GW data analysis, 
accounting for the detector's noise. 
A standard technique for the analysis of ringdown GWs 
is the frequency-domain approach 
(see, e.g., Ref.~\cite{Nakano:2018vay} for 
various methods to extract the ringdown GWs), and 
we can directly apply the orthonormal set of mode function
presented in Section~\ref{subsec:ON-fit}
with the (so named) noise-weighted inner product
in the matched filtering method~(see, e.g., 
Ref.~\cite{Uchikata:2020wsp}).
The implementation of this method to the GW data analysis
and the analysis of real data from GW detectors
will be a future work.

Another interesting future work would be to add BH's spin to our close-limit waveform. 
Although it is known for any field points that the $2.5$PN near-zone metric 
with spin effects~\cite{Tagoshi:2000zg,Faye:2006gx}, 
the numerical computation of either metric or curvature perturbations in Kerr spacetime 
is particularly challenging 
in the time domain~\cite{Sundararajan:2007jg,Dolan:2012jg,Long:2021ufh}, 
and would need much coding efforts. 
It is probably more viable, instead, to perturbatively include the spin effect 
into our non-spinning close-limit waveform (e.g.,~\cite{Lousto:2010qx}). 
It may be also possible to calculate the close-limit waveform in Kerr case 
making use of the standard frequency-domain technique (and taking the extreme mass-ratio limit). 
Because, in either case, the extension is involved, we leave them for future work.

From the theoretical point of view, 
it will be useful to include information 
on excitation coefficients and factors
for the fundamental tone and overtones of QNMs 
derived from the BH perturbation approach
(See, e.g., Refs.~\cite{Berti:2006wq,Zhang:2013ksa,Hughes:2019zmt,Lim:2019xrb}).
Also, although we focused only on the $(\ell=2,\,m=0)$ 
and $(\ell=3,\,m=1)$ modes in the bulk of this paper 
(and will focus only on the $(\ell=2,\,m=2)$ mode
in Appendix~\ref{app:fit_NR}), 
the inclusion of the other subdominant 
(higher harmonic) modes will be beneficial. 
In particular, the $(\ell=4,\,m=4)$ mode of NR waveforms
for binary BHs is especially interesting 
because this mode includes the second order modes composed 
by the non-linear, self-coupling of first-order 
$(\ell=2,\,m=2) \times (\ell=2,\,m=2)$ mode, 
and it has already been found
in NR simulations~\cite{London:2014cma}. 

Nevertheless, astute readers may ask 
what extent the results 
based on the close-limit approximation 
(to the head-on collisions of two nonspinning BHs) 
are consistent with the full NR waveforms 
in more astrophysically relevant scenarios 
(e.g., coalescences observed by LIGO/Virgo). 
We conclude our paper to briefly answer this question,   
analyzing the NR waveform of SXS:BBH:0305 in the 
Simulating eXtreme Spacetimes (SXS) catalogue~\cite{SXS} 
with the same approach that we took 
in Section~\ref{sec:results}; 
Appendices~\ref{app:fit_NR} and~\ref{app:fit-coeffs-NR} 
also provide supplemented examples of the analysis 
for NR waveforms.

Figure~\ref{fig:BBH0305-waveform} shows  
the $(\ell=2,\,m=2)$ spheroidal mode of the waveform 
SXS:BBH:0305~\cite{Lovelace:2016uwp,Boyle:2019kee,SXS}
(see Eq.~\eqref{Y-to-S} for the construction of the spheroidal modes from the spherical ones).
Here, we set the time of the peak amplitude 
as the starting time of the QNM fit 
(see the orange dashed line in the left panel of
Figure~\ref{fig:BBH0305-waveform}),  
and we use the QNM fitting formula introduced 
in Eq.~\eqref{eq:fit-formula-QNM-ON}.
Unlike the TD data generated within the close-limit approximation 
and the BH perturbation approach, 
we cannot confirm the power-law tail 
certainty due to some constant fit residuals
of $\sim 10^{-5}$ after $t-t_\textrm{peak} \gtrsim 80M$; 
recall that the tail contribution in Figure~\ref{fig:headon-waveform_2} is 
at the level of $\lesssim O(10^{-5})$, 
which is as large as the remaining residual in the NR waveform.
We expect that future NR simulation 
with higher numerical accuracy 
(or, e.g., the close-limit approximation 
to bound-orbit BBH mergers)
will provide a more robust answers 
of the need of the tail contribution
in the ringdown waveform modeling, and eventually 
in the measurement of remnant properties of BHs. 

Figure~\ref{fig:coeffs-convergence-NR} displays 
the fitting coefficients 
for the original QNM basis $C_k^{\ell m}$
and those for the orthonormalized QNM basis 
$\tilde{C}_k^{\ell m}$ 
introduced in Eq.~\eqref{eq:mode-coeffs}, 
again focusing on the the $(\ell=2,\,m=2)$ 
spheroidal mode of SXS:BBH:0305.
We see that the trend of $C_k^{\ell m}$ and $\tilde{C}_k^{\ell m}$ are similar to 
that of the close-limit, TD data 
displayed in Figures~\ref{fig:coeffs-convergence}
and~\ref{fig:coeffs-convergenceB}
(see also Appendix~\ref{app:fit-coeffs-NR}).
Namely, $\tilde{C}_k^{\ell m}$ converges 
as increasing the overtone $k$ 
while $C_k^{\ell m}$ (from the least square fit 
by using Eq.~\eqref{eq:fit-formula-QNM}) does not.
The convergence property of $\tilde{C}_k^{\ell m}$ 
supports the argument 
that the dominant $(\ell=2,\,m=2)$ mode of NR waveforms 
after the time of the peak amplitude can be described only 
by the QNMs in the linearized BH perturbation  
(see Section~\ref{sec:introduction} for the other confirmations).

\vspace{6pt} 



\authorcontributions{
The authors contribute equally to this paper.
}

\funding{
N.~S. and H.~N. acknowledge support from JSPS KAKENHI Grant No. JP21H01082 and No. JP17H06358.
S.~I. acknowledges support from STFC through Grant No. ST/R00045X/1. 
S.~I. also thanks to networking support by the GWverse COST Action CA16104, 
``Black holes, gravitational waves and fundamental physics.''
H.~N. acknowledges support from JSPS KAKENHI Grant No. JP21K03582.
}

\acknowledgments{
We would like to thank Ryuichi Fujita and Takahiro Tanaka for useful discussion. 
S.~I. is grateful to Leor Barack for his continuous encouragement.
}

\conflictsofinterest{The authors declare no conflict of interest.} 



\appendixtitles{yes} 
\appendix

\section{Analysis of some numerical-relativity waveforms in time domain} 
\label{app:fit_NR}

In this appendix and Appendix~\ref{app:fit-coeffs-NR}, 
we focus on the $(\ell=2,\,m=2)$ (spheroidal harmonic) 
mode of the ringdown signal 
(after the time of the peak amplitude) provided 
by NR waveforms for nonprecessing, 
spinning binary BH mergers, 
and analyze them following the same approach as in Section~\ref{sec:results}. 
Our goal here is to examine if we find i) the late-time, power-law tail
(this appendix) 
and ii) the convergence of the orthonormalized QNM fits 
(Appendix~\ref{app:fit-coeffs-NR}) 
in more astrophysically relevant NR waveforms.
We note that there are also various works on the mismatch and parameter estimation errors 
based on the ringdown portion of the NR waveforms and their QNM fits 
with overtones~\cite{Giesler:2019uxc,Bhagwat:2019dtm,Ota:2019bzl,Cook:2020otn,JimenezForteza:2020cve,Forteza:2021wfq},
memory (mainly the $m=0$ mode)~\cite{Mitman:2020pbt}
and mirror (negative $m$) modes~\cite{Dhani:2020nik,Dhani:2021vac}.

\begin{table}[ht]
\caption{Masses and spins of remnant BHs from NR simulations of binary BHs 
examined in this Appendix. $M_{\rm rem}/M$ and $\chi_{\rm rem}$
are the nondimensional mass ans spin parameters, respectively. 
The `SXS' data is presented in 
the Simulating eXtreme Spacetimes (SXS) catalogue~\cite{SXS} 
while the `RIT' data is imported from  
the CCRG@RIT Catalog of Numerical Simulations~\cite{RIT}.
Each reference in the table is cited from the corresponding metadata file.
SXS:BBH:0305 is used in Section~\ref{sec:discussion},
and the remaining NR data are studied in Appendices~\ref{app:fit_NR}
and~\ref{app:fit-coeffs-NR}.}
\label{table:remnant_BHs}
\centering
\begin{tabular}{lccc}
\toprule
ID 
& Mass ($M_{\rm rem}/M$) & Spin ($\chi_{\rm rem}$) &
Reference
\\
\midrule
SXS:BBH:0305 & 0.952032939704 & 0.6920851868180025
& \cite{Lovelace:2016uwp,Boyle:2019kee,SXS}
\\
\midrule
SXS:BBH:1936 & 0.9851822160611967 & 0.021659378750190413
& \cite{Varma:2019csw,Boyle:2019kee,SXS}
\\
SXS:BBH:0260 & 0.9810057011067479 & 0.12447236057508855
& \cite{Chu:2015kft,Boyle:2019kee,SXS}
\\
SXS:BBH:1501 & 0.93633431069 & 0.8085731624240002
& \cite{Varma:2018mmi,Boyle:2019kee,SXS}
\\
SXS:BBH:1477 & 0.911077401717 & 0.907542632208
& \cite{Varma:2018mmi,Boyle:2019kee,SXS}
\\
SXS:BBH:0178 & 0.8866898235070239 & 0.9499311295284206
& \cite{Scheel:2014ina,Boyle:2019kee,SXS}
\\
SXS:BBH:1124 & 0.8827804590335694 & 0.9506671398803149
& \cite{Boyle:2019kee,SXS}
\\
\midrule
RIT:BBH:0062 & 0.9520211506 & 0.6919694604
& \cite{Healy:2017psd,Healy:2019jyf,Healy:2020vre}
\\
RIT:BBH:0604 & 0.9361520656 & 0.8101416903
& \cite{Healy:2017psd,Healy:2019jyf,Healy:2020vre}
\\
RIT:BBH:0558 & 0.9108618514 & 0.9077062488
& \cite{Healy:2017psd,Healy:2019jyf,Healy:2020vre}
\\
RIT:BBH:0767 & 0.9057246958 & 0.9462438132
& \cite{Healy:2017psd,Healy:2019jyf,Healy:2020vre}
\\
\bottomrule
\end{tabular}
\end{table}

In our analysis of the NR waveforms, we fix the values of remnant BH mass
$M_\textrm{rem}/M$ and spin $\chi_\textrm{rem}$ 
provided by the NR simulations 
(see Table~\ref{table:remnant_BHs})
and calculate the QNM frequencies 
by assuming a Kerr geometry with these remnant properties.
We also mean the $(\ell=2,\,m=2)$ mode 
as that in terms of the spin-weighted spheroidal harmonics. 
Because the NR waveforms rather adopt the spherical basis, 
we have to account for the mixing of spheroidal 
and spherical bases here.
The GW strain is decomposed in terms of the spherical and spheroidal 
harmonics as
\begin{align}\label{Y-to-S}
h =& \frac{1}{r} \sum_{\ell,\,m} h^{\ell m}(t)
{}_{-2}P_{\ell m}(\theta)
\frac{e^{im\varphi}}{\sqrt{2\pi}} \cr
=& \frac{1}{r} \sum_{\ell',\,m} \tilde{h}^{\ell' m}(t)
{}_{-2}S_{\ell' m}^{a\omega}(\theta)
\frac{e^{im\varphi}}{\sqrt{2\pi}} \,,
\end{align}
where ${}_{-2}P_{\ell m}(\theta)$ and 
${}_{-2}S_{\ell m}^{a\omega}(\theta)$
are the spin-weighted associated Legendre function and 
the spin-weighted spheroidal harmonics, respectively.
When calculating the $(\ell=2,\,m=2)$ spheroidal harmonic mode,
we take into account the mixing of 
$2\le\ell\le 8$ (for the SXS data) and $2\le\ell\le 4$ 
(for the RIT data) spherical harmonic modes by using the
relation as
\begin{equation}
\tilde{h}^{\ell' m}(t) =
\int_0^{\pi} d\theta \, \sin\theta \, \sum_{\ell}
h^{\ell m}(t) {}_{-2}P_{\ell m}(\theta)
{}_{-2}S_{\ell' m}^{a\omega}(\theta) \,.
\label{eq:strain_spheroidal}
\end{equation}

\begin{figure}[ht]
\centering
\includegraphics[width=0.9\textwidth]{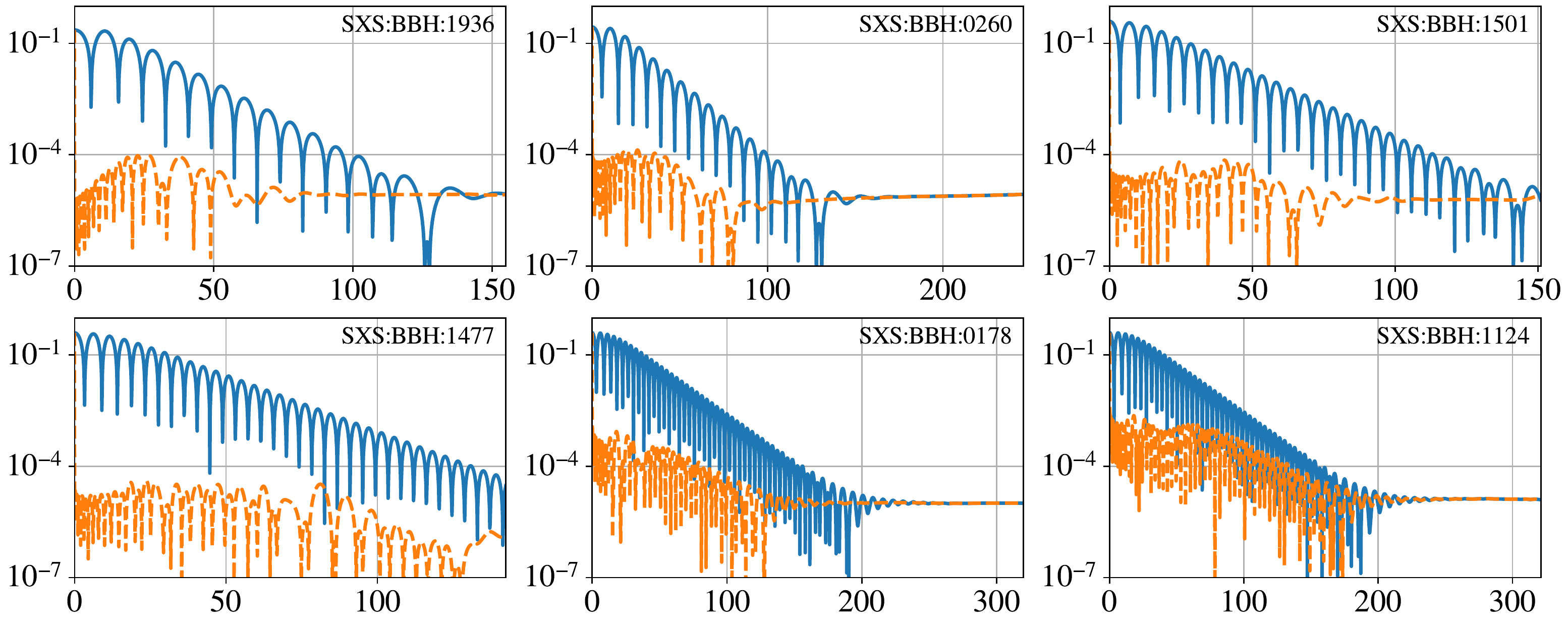}
\caption{Real part of the $(\ell=2,\,m=2)$ 
spheroidal harmonic mode $\Re(\tilde{h}^{22})$
of SXS NR waveforms 
listed in Table~\ref{table:remnant_BHs}
and their QNM fit residuals. Each plot corresponds to
that in the right panel of Figure~\ref{fig:BBH0305-waveform} 
in the case of SXS:BBH:0305.
}
\label{fig:SXS-residue}
\end{figure}

\begin{figure}[ht]
\centering
\includegraphics[width=0.6\textwidth]{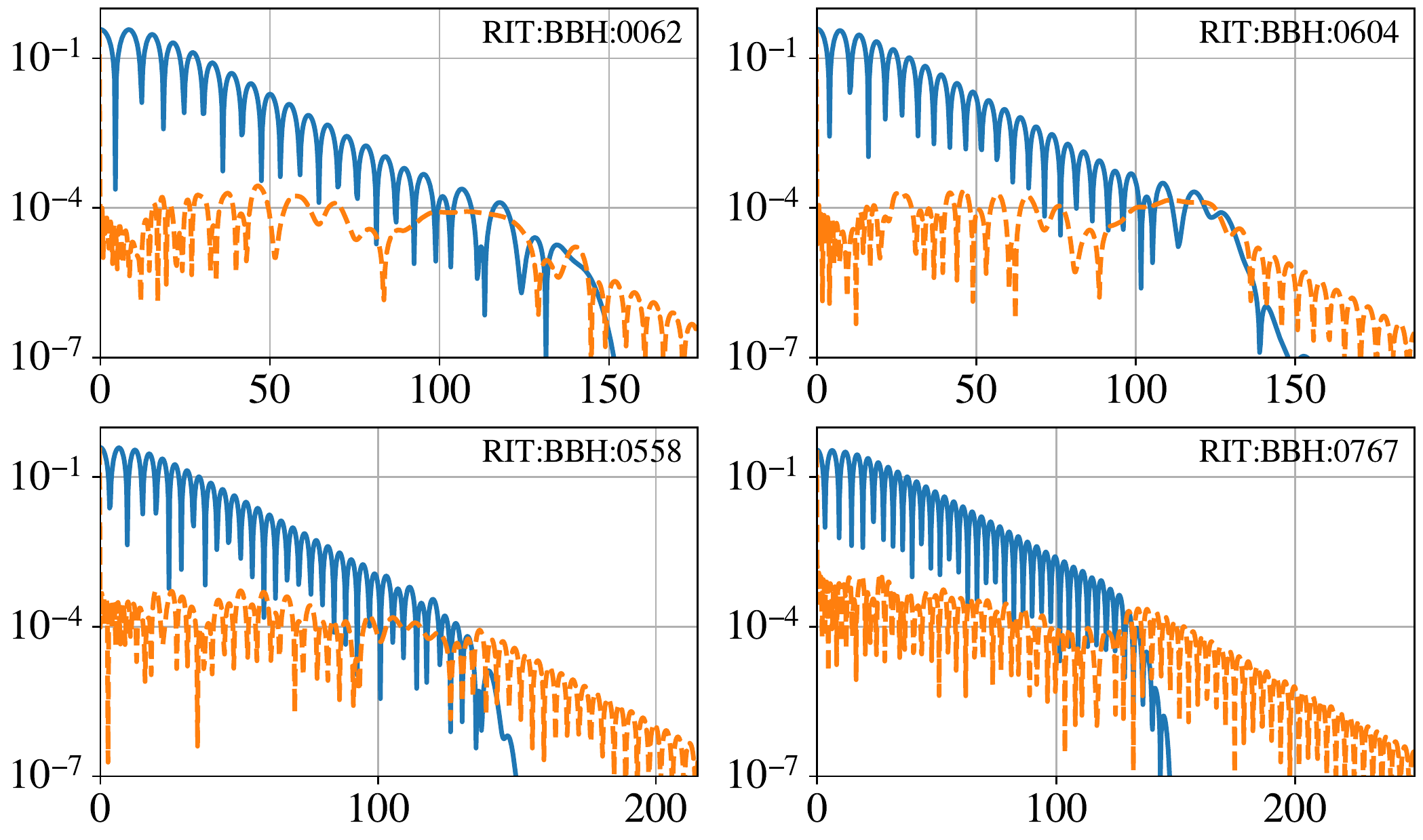}
\caption{Real part of the $(\ell=2,\,m=2)$ 
spheroidal harmonic mode $\Re(\tilde{h}^{22})$
of RIT NR waveforms 
listed in Table~\ref{table:remnant_BHs}
and their QNM fit residuals. 
}
\label{fig:RIT-residue}
\end{figure}

Figure~\ref{fig:SXS-residue} shows the QNM fit residuals of 
the $(\ell=2,\,m=2)$ SXS ringdown waveforms: 
SXS:BBH:1936 (almost nonspinning remnant BH, $\chi_{\rm rem} \sim 0.02$)~\cite{Varma:2019csw,Boyle:2019kee,SXS},
SXS:BBH:0260 (slowly rotating remnant BH, $\chi_{\rm rem} \sim 0.12$)~\cite{Chu:2015kft,Boyle:2019kee,SXS},
SXS:BBH:1501 (spinning remnant BH, $\chi_{\rm rem} \sim 0.81$)~\cite{Varma:2018mmi,Boyle:2019kee,SXS},
SXS:BBH:1477 (spinning remnant BH, $\chi_{\rm rem} \sim 0.91$)~\cite{Varma:2018mmi,Boyle:2019kee,SXS},
SXS:BBH:0178 (highly spinning remnant BH, $\chi_{\rm rem} \sim 0.95$)~\cite{Scheel:2014ina,Boyle:2019kee,SXS},
and SXS:BBH:1124 (highly spinning remnant BH, $\chi_{\rm rem} \sim 0.95$)~\cite{Boyle:2019kee,SXS}.
Here, we use only $\Re(\tilde{h}^{\ell m})$ 
from Eq.~\eqref{eq:strain_spheroidal} of the NR waveforms
and apply the real expression for the QNM fit model
in Eq.~\eqref{eq:fit-formula-QNM-ON} with overtones 
up to $N = 7$ 
(by replacing $\Psi^{\ell m}$ 
with $\Re(\tilde{h}^{\ell m})$
and accounting for the orthonormalization of QNM basis 
in Section~\ref{subsec:ON-fit}).

We find that all the SXS waveform data in the late time have 
the same constant residue of $O(10^{-5})$ as for SXS:BBH:0305, 
except for BBH:1477 that has the shorter time stretch between the peak 
and the end of data including the original data than other data set. 
We cannot identify the late-time tails in the QNM fit 
residuals (at least) in these NR waveforms, 
and some improvement in accuracy of NR simulation 
may be needed in order to reveal the tails.

We also find the slower damping fit residuals 
than the $(\ell=2,\,m=2)$ mode itself
in the late-time part of SXS:BBH:0178 and SXS:BBH:1124 
(with highly spinning remnant BHs, $\chi_\textrm{rem} \sim 0.95$). 
This is rather unexpected result because any QNMs 
(even including the second and higher order perturbations) 
have a faster damping time than the fundamental tone ($\ell=2,\,m=2,\,n=0$) 
in the high-spin range; see Figure~\ref{fig:QNM_0999} 
in Appendix~\ref{app:qnm_freq}. 

To have additional evidence for the above 
slower damping fit residuals, 
we examine another NR waveform set produced 
by RIT team~\cite{Healy:2017psd,Healy:2019jyf,Healy:2020vre}, including 
RIT:BBH:0062
(spinning remnant BH, $\chi_{\rm rem} \sim 0.69$),
RIT:BBH:0604
(spinning remnant BH, $\chi_{\rm rem} \sim 0.81$),
RIT:BBH:0558
(spinning remnant BH, $\chi_{\rm rem} \sim 0.91$),
and RIT:BBH:0767
(highly spinning remnant BH, $\chi_{\rm rem} \sim 0.95$), 
and display their QNM fit residuals in Figure~\ref{fig:RIT-residue}. 
RIT:BBH:0767 (bottom-right) has the remnant spin as high as 
that of SXS:BBH:0178 and SXS:BBH:1124, 
but we cannot confidently identify the similar slow-damping residuals 
due to the larger numerical error at $\sim 10^{-4}$. 
The source of this slower damping residuals remains 
unclear to us, 
and more systematic investigation awaits future work.

Before concluding this appendix, we should note the possible 
second-order contribution, i.e., the self-coupling 
of the two first-order QNMs computed from the linear 
perturbation theory, to our analysis. 
For example, the second-order, $(\ell=2,\,m=2) \times (\ell=2,\,m=2)$ mode 
is found in NR simulations~\cite{London:2014cma}.  
This mode is corresponding to the second order $(\ell=4,\,m=4)$ mode, 
and we did not consider it here. 
Also, in principle, the $(\ell=2,\,m=2)$ mode include the second-order contribution 
like $(\ell=2,\,m=1) \times (\ell=2,\,m=1)$ mode, 
but we cannot find these second-order contributions 
in our analysis. 
This is consistent with the previous work in Ref.~\cite{London:2014cma}.

\section{Analysis of the fitting coefficients of some numerical-relativity data}
\label{app:fit-coeffs-NR}

\begin{figure}[ht]
\centering
\includegraphics[width=0.9\textwidth]{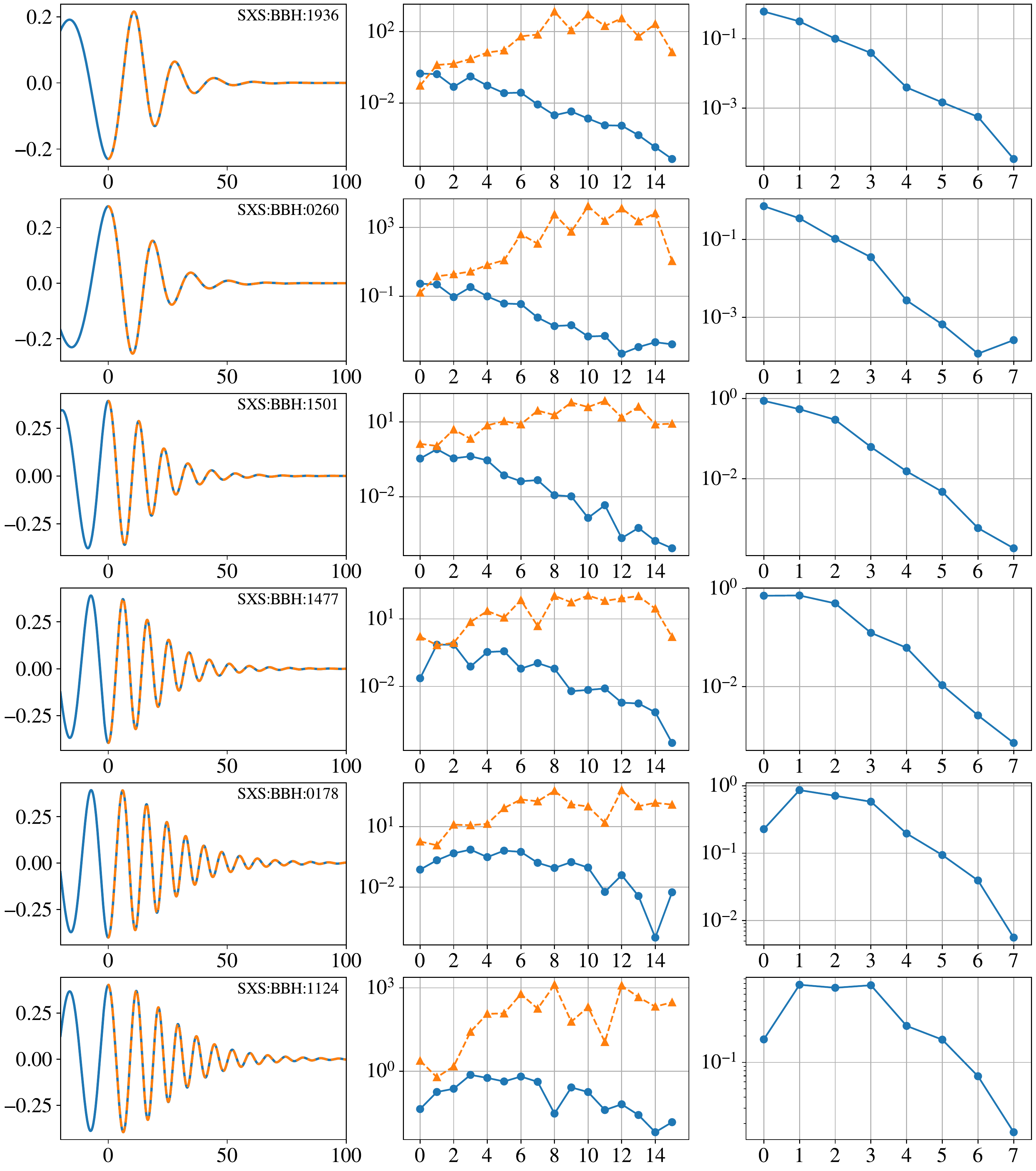}
\caption{The waveforms and fitting coefficients of samples
of SXS data listed in Table~\ref{table:remnant_BHs}.
The plots in the left column show the real part of $(\ell=2,\,m=2)$ spheroidal harmonic mode (the blue solid line)
and the QNM fit to the model with the orthonormal set given in Eq.~\eqref{eq:fit-formula-QNM-ON}
(the orange dashed line). The plots in the center column present the absolute values of the
fitting coefficients for $\tilde{C}_k^{\ell m}$ 
(the blue filled circles) and $C_k^{\ell m}$ (the orange filled triangles).
The plots in the right column show $\Delta_n$ in Eq.~\eqref{eq:Delta_n}.}
\label{fig:SXS-example}
\end{figure}

\begin{figure}[ht]
\centering
\includegraphics[width=0.9\textwidth]{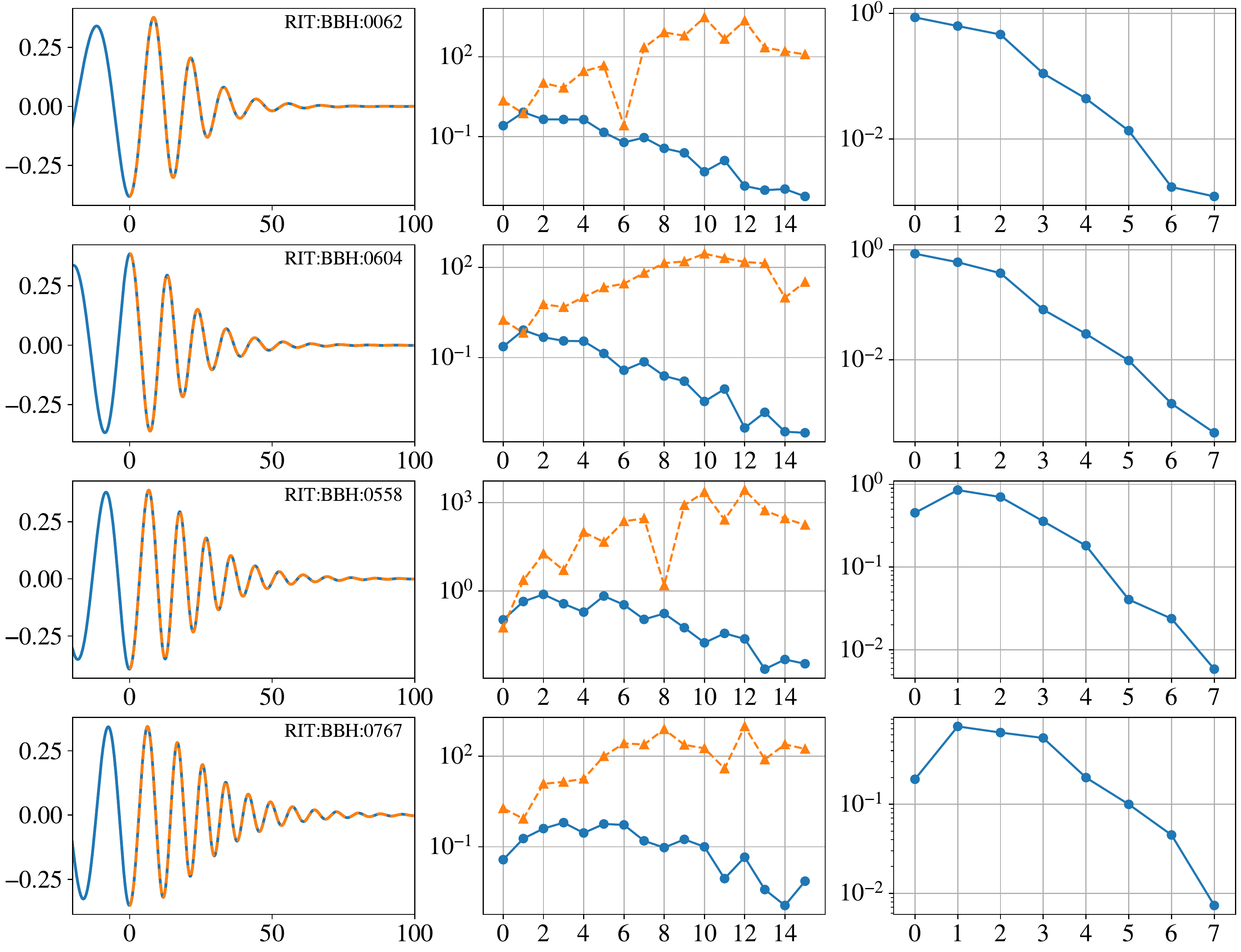}
\caption{The waveforms and fitting coefficients of samples
of RIT data listed in Table~\ref{table:remnant_BHs}.
The plots in the left column show the real part of the $(\ell=2,\,m=2)$ 
spheroidal harmonic mode (the blue solid line)
and the QNM fit to the model with the orthonormal set given in Eq.~\eqref{eq:fit-formula-QNM-ON}
(the orange dashed line). The plots in the center column present the absolute values of the
fitting coefficients for $\tilde{C}_k^{\ell m}$ 
(the blue filled circles) and $C_k^{\ell m}$ (the orange filled triangles).
The plots in the right column show $\Delta_n$ in Eq.~\eqref{eq:Delta_n}.}
\label{fig:RIT-example}
\end{figure}

As a by-product of the analysis in Appendix~\ref{app:fit_NR}, we obtain
the fitting coefficients (in terms of 
the original QNM basis
and the orthonormal set of mode functions)
for the NR data listed in Table~\ref{table:remnant_BHs}, 
and can check the convergence in the same manner as shown in 
Figure~\ref{fig:coeffs-convergence-NR}. We summarize the results in
Figure~\ref{fig:SXS-example} for the SXS data and 
Figure~\ref{fig:RIT-example} for the RIT data.
From these results, we expect that the contribution 
from the overtones to the ringdown waveform 
becomes more significant with larger remnant spins.
The further study will be required to clarify the relation 
between the parameters of binary BHs and the significance of the overtones
(A comprehensive study of the dependence of the mode excitations on the
source parameters has been done in the extreme mass-ratio limit~\cite{Hughes:2019zmt,Lim:2019xrb}).

\section{Frequencies of Kerr quasinormal modes}
\label{app:qnm_freq}

Although many plots for Kerr QNM frequencies are available 
in literature, e.g., 
in Refs.~\cite{Onozawa:1996ux,Berti:2009kk,Cook:2014cta},
we reproduce a QNM figure here for $m=\ell$ modes 
using publicly available data provided 
in ``Ringdown'' by Berti~\cite{QNM_EB}
(see also ``Kerr Quasinormal Modes: $s=-2$, $n=0$--$7$''
by Cook~\cite{QNM_GC}). 
They give the lowest damping rate in each $\ell$ mode, 
and help to check the presence of 
the second order perturbation in our analysis. 
Below the indices $(\ell,\,m)$ refer 
to the spin-weighted spheroidal basis with $s = -2$.

\begin{figure}[ht]
\centering
\includegraphics[width=0.9\textwidth]{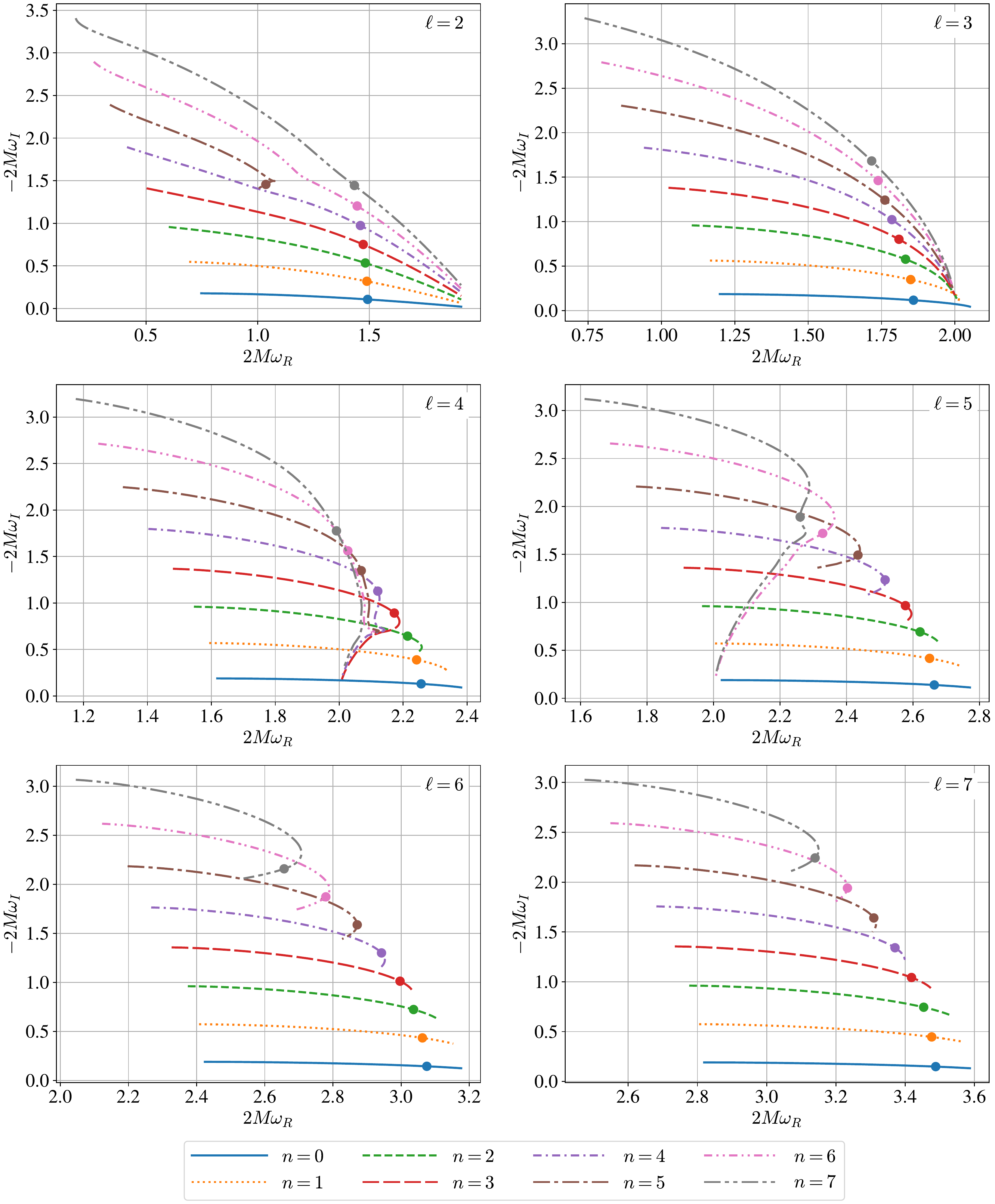}
\caption{
Fundamental tone and overtones of Kerr QNMs for $m=\ell$ modes. 
Here $\omega_R$ and $\omega_I$ mean the real and imaginary parts
of the QNM frequency, respectively.
We plot the values for $2 \le \ell \le 7$
in the range of the nondimensional spin of
$0\le \chi \le 0.999$.
The filled circles show the QNM values at $\chi=0.95$.
For $\chi \le 0.95$, the $n=0$ mode
always has the highest real frequency
and the slowest damping time.
There is an interesting feature in the fifth overtone ($n=5$)
of the $m=\ell=2$ mode.
To make these plots, we use the QNM data provided in
Ref.~\cite{QNM_EB}.}
\label{fig:QNM_0999}
\end{figure}

Figure~\ref{fig:QNM_0999} shows the Kerr QNM frequencies
(the fundamental tone and overtones up to $n=7$)
for the nondimensional spin $0 \le \chi \le 0.999$, 
but it suffices to consider the QNM frequencies 
up to the filled circles ($\chi = 0.95$) 
because the NR simulations used for our study 
have the remnant BH spin $\chi \lesssim 0.95$ after merger. 
In particular, we see that the $(\ell=2,\,m=2,\,n=0)$ mode 
has the lowest damping rate (recall Eq.~\eqref{eq:def-tau}).
Importantly, the first order $(\ell=2,\,m=2,\,n=0)$ mode remains 
to be the longest-lived mode even when considering 
the second and higher order QNMs as we establish below.

Suppose an expansion of the metric of the form,
\begin{equation}
g_{\mu\nu} = g_{\mu\nu}^{(0)} + h_{\mu\nu}^{(1)}
+ h_{\mu\nu}^{(2)} + (\text{higher-order terms}) \,,
\end{equation}
where $g_{\mu\nu}^{(0)}$ is a background (Kerr) spacetime,
and $h_{\mu\nu}^{(1)}$ is the first order perturbation here.
The second order perturbation $h_{\mu\nu}^{(2)}$ formally satisfies 
the linearized Einstein equation 
\begin{equation}
G_{\mu\nu}^{(1)}[h^{(2)}]
= - G_{\mu\nu}^{(2)}[h^{(1)},\, h^{(1)}] \,,
\end{equation}
where $G_{\mu\nu}^{(1)}$  is the linearized Einstein
tensor (in a given gauge), and 
$G_{\mu\nu}^{(2)}[h^{(1)},\, h^{(1)}]$ is the term
quadratic in $h^{(1)}$ in the expansion of the Einstein tensor. 
Although the real part of second-order QNM frequencies of $h_{\mu\nu}^{(2)}$ 
can take various values due to the self-mode coupling of $h_{\mu\nu}^{(1)}$ 
through $G_{\mu\nu}^{(2)}$, 
Figure~\ref{fig:QNM_0999} indicates that for 
$\chi \lesssim 0.95$ 
the imaginary part of any second-order QNM frequencies $\omega_I^{(2)}$ 
is bounded at~\cite{Ioka:2007ak,Nakano:2007cj}
\begin{equation}
\left|\omega_I^{(2)}\right| 
\geq 2\,\left|\omega_{I,\,220}^{(1)}\right| \,,
\end{equation}
where $\omega_{I,\,220}^{(1)}$ is
the imaginary part of the first order $(\ell=2,\,m=2,\,n=0)$ QNM frequency. 
This in turn implies that 
the damping time of any second order QNMs satisfies
\begin{equation}
\tau^{(2)} \leq \frac{1}{2}\,\tau_{220} \,.
\end{equation}
Therefore, the first order $(\ell=2,\,m=2,\,n=0)$ QNM has the slowest 
damping time, at least to the second order 
in the expansion of the metric  
and for $0 \le \chi \le 0.95$ of the remnant BH 
(whose imaginary part of the first order QNMs 
is not too close to zero).

\reftitle{References}



\externalbibliography{yes}

\bibliography{references}



\end{document}